\documentclass[useAMS,usenatbib]{mn2e}

\usepackage{mathtools}
\usepackage{amsmath}
\usepackage{amssymb}
\usepackage{graphicx}
\usepackage{xcolor}
\usepackage{hyperref} 
\usepackage[all]{hypcap}
\hypersetup{linkbordercolor=0 1 0}

\title[H$\alpha$ EW at z $\sim$ 5 from IRAC Colors]{Mean H$\alpha$+[NII]+[SII] EW Inferred for Star-Forming Galaxies at z $\sim$ 5.1-5.4 Using High-Quality Spitzer/IRAC Photometry}
\author[Rasappu et al.]{N. Rasappu$^{1}$, R. Smit$^{2}$, I. Labb\'{e}$^{1}$, R.J. Bouwens$^{1}$, D.P. Stark$^{3}$, R.S. Ellis$^{4}$, P.A. Oesch$^{5}$\\
$^{1}$Leiden Observatory, Leiden University, NL-2300 RA Leiden, The Netherlands\\
$^{2}$Department of Physics, Durham University, South Road, Durham DH1 3LE, UK\\
$^{3}$Department of Astronomy/Steward Observatory, 933 North Cherry Avenue, Tucson, AZ 85721\\
$^{4}$Cahill Center for Astronomy and Astrophysics, California Institute of Technology, MC 249-17, Pasadena, CA 91125, USA\\
$^{5}$Department of Astronomy, Yale University, New Haven, CT 06520}
\begin{document}

\date{Accepted ... Received 2014 December ...; in original form ...}

\pagerange{\pageref{firstpage}--\pageref{lastpage}} \pubyear{2015}

\maketitle

\label{firstpage}

\begin{abstract}
Recent $Spitzer$/InfraRed Array Camera (IRAC) photometric observations
have revealed that rest-frame optical emission lines contribute
signficantly to the broadband fluxes of high-redshift galaxies.
Specifically, in the narrow redshift range $z$ $\sim$ 5.1 - 5.4 the
{[3.6]} $-$ {[4.5]} color is expected to be very red, due to
contamination of the 4.5 $\mu$m band by the dominant H$\alpha$ line,
while the 3.6 $\mu$m filter is free of nebular emission lines.
We take advantage of new reductions of deep $Spitzer$/IRAC imaging
over the GOODS-North+South fields (\hyperlink{Labbe2015}{Labb\'e et
  al. 2015}) to obtain a clean measurement of the mean H$\alpha$
equivalent width from the {[3.6]} $-$ {[4.5]} color in the redshift
range $z$ = 5.1 - 5.4.
The selected sources either have measured spectroscopic redshifts (13
sources) or lie very confidently in the redshift range $z$ = 5.1-5.4
based on the photometric redshift likelihood intervals (11 sources).
Our $z_{phot}$ = 5.1-5.4 sample and $z_{spec}$ = 5.10-5.40
spectroscopic sample have a mean {[3.6]} $-$ {[4.5]} color of 0.31
$\pm$ 0.05 mag and 0.35 $\pm$ 0.07 mag, implying a rest-frame
equivalent width EW(H$\alpha$+[NII]+[SII]) of 665 $\pm$ 53 \AA{} and
707 $\pm$ 74 \AA{}, respectively, for sources in these samples.  These
values are consistent albeit slightly higher than derived by Stark et
al. 2013 at $z\sim4$, suggesting an evolution to higher values of the
H$\alpha$+[NII]+[SII] EW at $z>2$.  Using the {3.6 $\mu$m} band, which
is free of emission line contamination, we perform robust SED fitting
and find a median specific star formation rate of sSFR =
$17_{-5}^{+2}$ Gyr$^{-1}$, $7_{-2}^{+1}\times$ higher than at
$z\sim2$.  We find no strong correlation ($<$2$\sigma$) between the
H$\alpha$+[NII]+[SII] EW and the stellar mass of sources.  Before the
advent of JWST, improvements in these results will come through an
expansion of current spectroscopic samples and deeper $Spitzer$/IRAC
measurements.
\end{abstract}


\section{Introduction}

In recent years, large multi-wavelength photometric surveys conducted
with the $Hubble$ and $Spitzer$ $Space$ $Telescopes$ have enabled us
to study the properties of galaxies over cosmic time. Synthetic
stellar population modeling of broadband spectral energy distributions
(SEDs) has led to the determination of various physical properties
(e.g., stellar mass, star formation rate, age, dust extinction) of
these galaxies (\hyperlink{Eyles2005}{Eyles et al.}
\hyperlink{Eyles2005}{2005}; \hyperlink{Yan2006}{Yan et al.}
\hyperlink{Yan2006}{2006}; \hyperlink{Stark2009}{Stark et al.}
\hyperlink{Stark2009}{2009}; \hyperlink{Gonzalez2010}{Gonzalez et al.}
\hyperlink{Gonzalez2010}{2010}; \hyperlink{Bouwens2009}{Bouwens et
  al.} \hyperlink{Bouwens2009}{2009}, \hyperlink{Bouwens2012b}{2012};
Finkelstein et al.\ 2012; \hyperlink{Oesch2013}{Oesch et al.}
\hyperlink{Oesch2013}{2013}; \hyperlink{debarros2014}{de Barros et
  al.} \hyperlink{debarros2014}{2014}; \hyperlink{Salmon2015}{Salmon
  et al.}  \hyperlink{Salmon2015}{2015}).  Of all derived quantities,
stellar masses are particularly robust in stellar population
fits. Small changes in the age of the stellar populations, metallicity
or other parameters have no significant effect on the estimated masses
(e.g., \hyperlink{Finlator2007}{Finlator et al.}
\hyperlink{Finlator2007}{2007}; \hyperlink{Yabe2009}{Yabe et al.}
\hyperlink{Yabe2009}{2009}).

Considerable progress has been made in refining current
characterization of galaxies from the observations.  Even so, the
measurements of the specific star formation rate (sSFR, the star
formation rate divided by the stellar mass) have presented a puzzle to
the theoretical understanding of the build-up of mass in
galaxies. (e.g., \hyperlink{Bouche2010}{Bouch\'{e} et al.}
\hyperlink{Bouche2010}{2010}; \hyperlink{Weinmann2011}{Weinmann et
  al.} \hyperlink{Weinmann2011}{2011}).  Several past studies had
indicated that the sSFR of sources with fixed stellar mass shows no
evidence for significant evolution between $z$ $\simeq$ 2 and $z$
$\simeq$ 7 (\hyperlink{Stark2009}{Stark et al.}
\hyperlink{Stark2009}{2009}; \hyperlink{Gonzalez2010}{Gonzalez et al.}
\hyperlink{Gonzalez2010}{2010}).  This result was in apparent
disagreement with semi-analytic models predicting a strong increase in
the specific inflow rate (i.e. inflow rate divided by halo mass) of
baryons with redshift (\hyperlink{NeisteinDekel2008}{Neistein \&
  Dekel} \hyperlink{NeisteinDekel2008}{2008}).

Subsequent work has strongly suggested that the sSFR of galaxies is
somewhat higher at $z>2$ than it is at $z\sim2$ (e.g., Schaerer \& de
Barros 2010; Stark et al.\ 2013; Gonzalez et al.\ 2014; Salmon et
al.\ 2015).  The change in the inferred sSFR evolution with cosmic
time was the result of improved observational constraints and a more
sophisticated treatment of those constraints.  One example of this is
in a consideration of dust extinction in computing the sSFRs at $z>2$.
While it was not possible to properly account for the impact of dust
extinction in initial work (e.g., \hyperlink{Gonzalez2010}{Gonzalez et
  al.}  \hyperlink{Gonzalez2010}{2010}) due to large uncertainties on
the $UV$ colors of $z>2$ galaxies, later work
(\hyperlink{Bouwens2012b}{Bouwens et al.}
(\hyperlink{Bouwens2012b}{2012}) was able to make use of new
measurements of the $UV$ continuum slopes to account for the impact of
dust extinction, finding $\sim2\times$ larger sSFRs at $z\geq 4$.
This suggested a $2\times$ evolution relative to the $z\sim2$ value
(\hyperlink{Reddy2012b}{Reddy et al.}  \hyperlink{Reddy2012b}{2012}),
but still leaving the sSFR at $z$ $>$ 4 approximately constant.

Even more important has been the increasing awareness of the impact of
rest-frame optical nebular emission lines (e.g., H$\alpha$, [OIII],
[OII]) on the broadband fluxes (e.g.,
\hyperlink{SchaererdeBarros2009}{Schaerer \& de Barros}
\hyperlink{SchaererdeBarros2009}{2009},
\hyperlink{SchaererdeBarros2010}{2010}, \hyperlink{Schenker2013}{Schenker
  et al.}  \hyperlink{Schenker2013}{2013}).  At high redshifts these
emission lines are shifted into the infrared, contaminating the IRAC
measurements of the stellar continuum.  Inferred stellar masses from
fitting of stellar population models will then be overestimated,
resulting in an underestimate of the sSFR.

Since these strong rest-frame optical lines are inaccessible to
spectroscopy beyond $z\sim2$-3, the strength of the nebular emission
lines has been estimated from the contamination of the $Spitzer$/IRAC
3.6$\mu$m and 4.5$\mu$m bands for galaxies at $z$ $>$
3. \hyperlink{Shim2011}{Shim et al.}  (\hyperlink{Shim2011}{2011})
show that galaxies in the range 3.8 $<$ $z$ $<$ 5.0 are considerably
brighter at 3.6 $\mu$m than expected from the stellar continuum alone
and argue that this excess is due to strong H$\alpha$ line
emission. \hyperlink{Stark2013}{Stark et al.}
(\hyperlink{Stark2013}{2013}) derive an H$\alpha$ equivalent width
(EW) distribution by comparing the [3.6] $-$ [4.5] color of
spectroscopically confirmed galaxies in the redshift range 3.8 $<$ $z$
$<$ 5.0, where the H$\alpha$ line lies in the 3.6 $\mu$m band, with an
uncontaminated control sample at 3.1 $<$ $z$ $<$ 3.6.  The results
indicate a possible trend towards higher H$\alpha$ EWs at higher
redshifts, which is extremely important for estimating the sSFR at $z$
$>$ 5 (see also \hyperlink{Labbe2013}{Labb\'e et al.}
\hyperlink{Labbe2013}{2013}; \hyperlink{debarros2014}{de Barros et
  al.} \hyperlink{debarros2014}{2014}; \hyperlink{Smit2014}{Smit et
  al.}  \hyperlink{Smit2014}{2014}; \hyperlink{Smit2015}{Smit et al.}
\hyperlink{Smit2015}{2015}).  Independent evidence for high-EW nebular
lines having a large impact on the broadband fluxes of $z>3$ galaxies
was obtained from early near-infrared, multi-object spectroscopic
campaigns (e.g., \hyperlink{Schenker2013}{Schenker et al.}
\hyperlink{Schenker2013}{2013}; \hyperlink{Holden2015}{Holden et
  al. 2015}).

The highest-redshift window providing us a largely clean measurement
of the H$\alpha$ EW is the redshift range $z\sim5.10$-5.40.  Here the
flux excess due to the redshifted nebular emission lines gives rise to
significantly redder [3.6] $-$ [4.5] colors over this range, making it
possible to quantify the EW of H$\alpha$ at $z>5.1$ in a similar way
to that possible in the redshift range $z=3.8$ to 5.0.  By examining
the [3.6] $-$ [4.5] color of galaxies from the
\hyperlink{Bouwens2015}{Bouwens et al.}
(\hyperlink{Bouwens2015}{2015}) catalog (see \S3.1) over the Great
Observatories Origins Deep Survey (GOODS)-North and South Fields
(\hyperlink{Giavalisco2004}{Giavalisco et al.\ 2004}) in this highest
redshift window $z$ $\sim$ 5.1-5.4 where H$\alpha$ can be cleanly
measured,\footnote{At $z\gtrsim5.5$, constraints on the $H\alpha$ EWs
  are also possible from Spitzer/IRAC observations, but would need to
  rely on the stacking the fluxes of $z\gtrsim5.5$ galaxies in the
  much less sensitive 5.8 $\mu$m band.} we can derive approximate
constraints on the H$\alpha$ flux and EW at the highest redshift
currently accessible at reasonable S/N with current facilities.  This
allows us to obtain better constraints on the evolution of the mean
H$\alpha$ EW and sSFR as a function of redshift.

This paper is structured as follows.  In \S2, we describe the
observational data sets and our photometric selection of sources in
the narrow redshift range. In \S3, we examine our selected sample of
galaxies. We describe the assumptions made in deriving the EWs and
sSFRs.  Finally, in \S4, we discuss our results and give a summary.
We refer to the HST F435W, F606W, F775W, F814W, F850LP, F105W, F125W,
and F160W bands as $B_{435}$, $V_{606}$, $i_{775}$, $I_{814}$,
$z_{850}$, $Y_{105}$, $J_{125}$ and $H_{160}$, respectively.  For
consistency with previous work, we adopt the concordance model with
$\Omega _{m}$ = 0.3, $\Omega _{\Lambda }$ = 0.7 and $H_{0}$ = 70 km
s$^{-1}$ Mpc$^{-1}$.  Throughout we assume a
\hyperlink{Salpeter1955}{Salpeter} (\hyperlink{Salpeter1955}{1955})
initial mass function (IMF) between 0.1$-$100 $M_{\odot}$. All
magnitudes are quoted in the AB photometric system
(\hyperlink{OkeGunn1983}{Oke \& Gunn} \hyperlink{OkeGunn1983}{1983}).

\section[]{Observations}


\subsection{Data}

\begin{figure*}
\centering
\includegraphics[scale=0.60,trim=20mm 5mm 0mm 0mm]{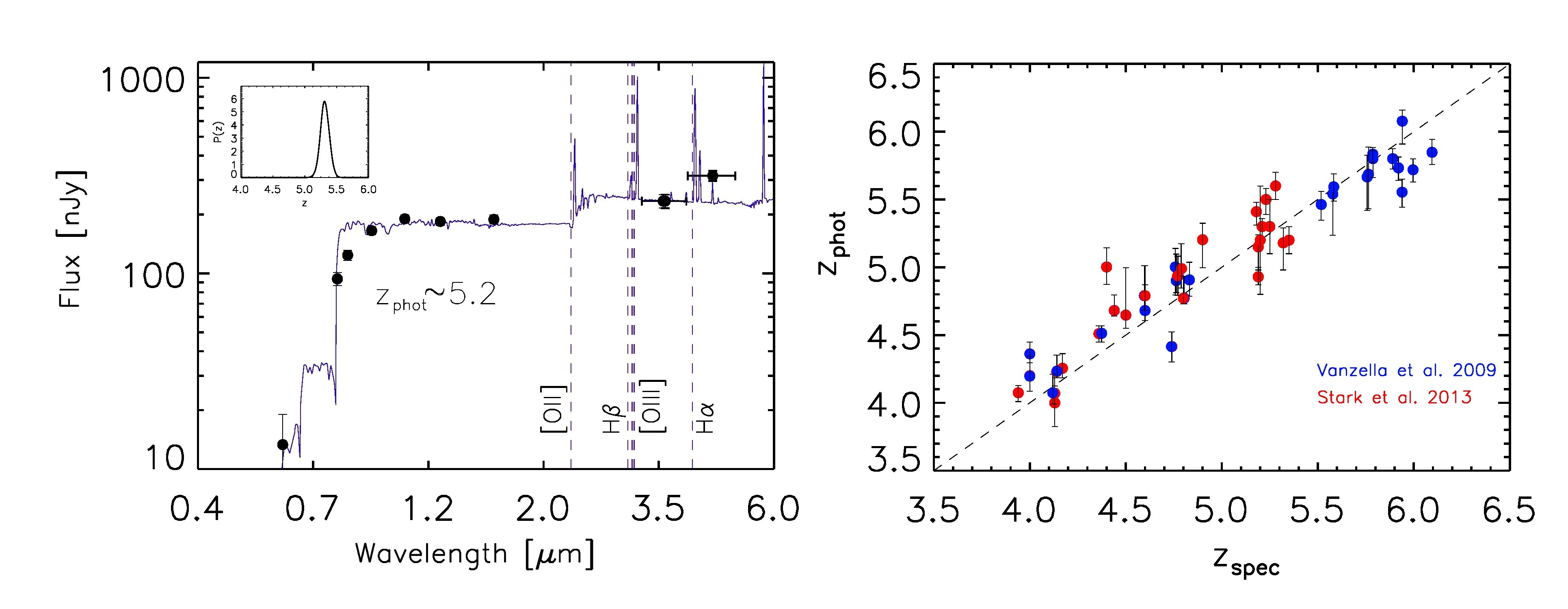}
\caption{\label{SED}$Left$ $panel$$:$  Model spectral energy distribution (blue) and the HST flux measurements (black dots) for one of the sources in our selection fitted with the photometric redshift code EAZY (\protect\hyperlink{Brammer2008}{Brammer et al.} \protect\hyperlink{Brammer2008}{2008}). Displayed in the upper left corner: the cumulative probability distribution P(z) to find the galaxies at a certain redshift. We only use candidates with $P(5.1<z<5.4)>0.85$. The $Spitzer$/IRAC bands are not included in the fitting to prevent bias in the [3.6] $-$ [4.5] colors. $Right$ $panel$$:$ Photometric redshift as determined by EAZY against spectroscopic redshift for the sources in \protect\hyperlink{Vanzella2009}{Vanzella et al.} (\protect\hyperlink{Vanzella2009}{2009}) and \protect\hyperlink{Stark2013}{Stark et al.} (\protect\hyperlink{Stark2013}{2013}). At $z$ $\lesssim$ 5, EAZY seems to overestimate the redshift by $\Delta z/(1+z)$ $\sim$ 0.02, while the redshift is underestimated by $\Delta z/(1+z)$ $\gtrsim$ 0.03 for $z$ $\gtrsim$ 6.0. Though we need a far larger sample to confirm the presence of such an offset, the redshift window that we are considering is seemingly the least affected.}
\label{fig:SED}
\end{figure*}

In order to select sources in the redshift range $z$ $\sim$ 5.1-5.4,
we make use of the deep optical/ACS and near-infrared/WFC3/IR
observations over the GOODS-North and GOODS-South Fields from three
significant $Hubble$ $Space$ $Telescope$ (HST) programs: GOODS, ERS
(Early Release Science: \hyperlink{Windhorst2011}{Windhorst et al.}
\hyperlink{Windhorst2011}{2011}), and CANDELS (Cosmic Assembly
Near-infrared Deep Extragalactic Legacy Survey:
\hyperlink{Grogin2011}{Grogin et al.}  \hyperlink{Grogin2011}{2011};
\hyperlink{Koekemoer2011}{Koekemoer et al.}
\hyperlink{Koekemoer2011}{2011}).  The moderately deep regions over
the CANDELS GOODS-North and South reach a 5$\sigma$ depth of $\sim$
27.5 mag in the $Y_{105}$, $J_{125}$ and $H_{160}$ filters with the
HST and cover $\sim$ 100 arcmin$^{2}$. The deep regions over the
CANDELS GOODS-North and South reach a 5$\sigma$ depth of $\sim$ 28.5
in the $Y_{105}$, $J_{125}$ and $H_{160}$ bands, covering $\sim$ 125
arcmin$^{2}$ (\hyperlink{Grogin2011}{Grogin et al.}
\hyperlink{Grogin2011}{2011}).  HST observations with the ACS
(Advanced Camera for Surveys) are available in the $B_{435}$,
$V_{606}$, $i_{775}$, $I_{814}$ and $z_{850}$ bands, up to $\sim$ 29
mag at 5$\sigma$ in $I_{814}$ (\hyperlink{Bouwens2015}{Bouwens et al.}
\hyperlink{Bouwens2015}{2015}).  Over the northern $\sim$40 arcmin$^2$
section of GOODS South (\hyperlink{Windhorst2011}{Windhorst et al.}
\hyperlink{Windhorst2011}{2011}), deep near-IR observations are
available ($\sim$28 mag at $5\sigma$) in the $Y_{098}$, $J_{125}$, and
$H_{160}$ bands and also in the $B_{435}$, $V_{606}$, $i_{775}$ and
$z_{850}$ bands with ACS. The observations are PSF matched to the
$H_{160}$ band before measuring the colors in scalable
\hyperlink{Kron1980}{Kron} (\hyperlink{Kron1980}{1980}) apertures.

Essential to the analysis we perform is the 3.6 $\mu$m and 4.5 $\mu$m
IRAC observations from the $Spitzer$ $Space$ $Telescope$.
$Spitzer$/IRAC data is from the original GOODS program, the Spitzer
Extented Deep Survey (SEDS: \hyperlink{Ashby2013}{Ashby et al.}
\hyperlink{Ashby2013}{2013}), the Spitzer Very Deep Survey (S-CANDELS:
\hyperlink{Ashby2015}{Ashby et al.}  \hyperlink{Ashby2015}{2015})
Exploration Science Project, the IUDF10 program
(\hyperlink{Labbe2015}{Labb\'e et al. 2015}), and other programs (such
as PID10076, PI: Oesch).  The Spitzer/IRAC reductions we utilize were
generated by \hyperlink{Labbe2015}{Labb\'e et al. (2015)} and feature
a 1.8$''$-diameter FWHM for the PSF.

Deblending neighboring galaxies in the IRAC observations and PSF
correctons are performed using the \textsc{mophongo} software
(\hyperlink{Labbe2010a}{Labb\'{e} et al.}
\hyperlink{Labbe2010a}{2010a}, \hyperlink{Labbe2010b}{2010b},
\hyperlink{Labbe2013}{2013}, \hyperlink{Labbe2015}{2015}). HST F160W images are used as a high
resolution prior to construct a model for the contaminating sources,
while leaving the normalization of the sources as a free parameter.
The fluxes of all sources in a radius of 13\arcsec \mbox{} are then
simultaneously fit to best match the IRAC image. Photometry is then
performed within a 2.0\arcsec \mbox{} diameter aperture. The deep IRAC
imaging from S-CANDELS reaches $\sim$26.8 mag at 5$\sigma$ in the
3.6 $\mu$m band.

For $z=5.1$-5.4 sources not included in the photometric catalogs of
Bouwens et al.\ (2015: due to their being located in areas of the
GOODS fields without $B_{435}$ or $Y_{098}$/$Y_{105}$-band
observations), we made use of the HST photometry from the 3D-HST
GOODS-North or GOODS-South catalogs (\hyperlink{Skelton2014}{Skelton
  et al.}  \hyperlink{Skelton2014}{2014}).  This is relevant for 5 $z$
= 5.1-5.4 galaxies in our final sample.  We refer to
\hyperlink{Skelton2014}{Skelton et al.}
(\hyperlink{Skelton2014}{2014}) for a detailed description.

\subsection{Photometric Redshift Selection}

\begin{figure}
\includegraphics[scale=0.7,trim=6mm 5mm 0mm 0mm]{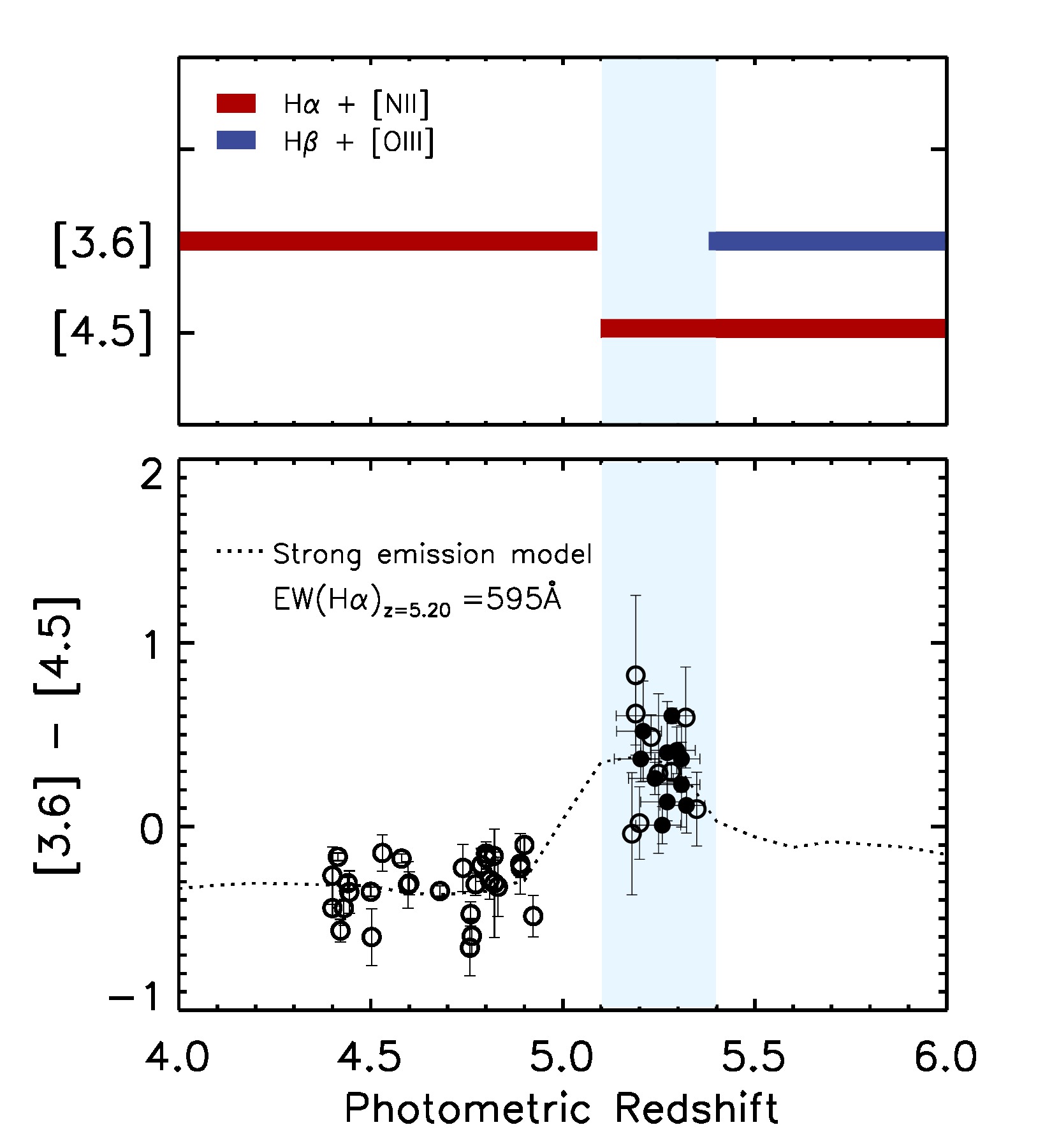}
\vspace{0em}
\caption{\label{nebularlines}Nebular emission line contamination in
  the 3.6 $\mu$m and 4.5 $\mu$m photometric filters. $Top$ $panel:$
  The redshift ranges over which the dominant nebular emission lines,
  H$\alpha$, H$\beta$, [OIII], [NII], and [SII], contribute to the 3.6
  $\mu$m and 4.5 $\mu$m flux measurements. $Lower$ $panel:$ The
  predicted [3.6] $-$ [4.5] color due to various nebular emission
  lines as a function of redshift. The solid black circles indicate
  sources that are selected in the redshift range $z_{phot}$ =
  5.1-5.4, where H$\alpha$ lies in the 4.5 $\mu$m filter, while 3.6
  $\mu$m is devoid of strong nebular emission. Their observed [3.6]
  $-$ [4.5] colors are primarily very red. The open circles are the
  colors for the spectroscopic sample found in
  \protect\hyperlink{Vanzella2009}{Vanzella et al.}
  (\protect\hyperlink{Vanzella2009}{2009}), which we use to estimate a
  stellar continuum color of $\sim$ 0.00 mag. The dotted line
  indicates the expected color for a strong evolution of the
  rest-frame equivalent width, EW$_{0}$(H$\alpha$) $\propto$
  (1+$z$)$^{1.8}$ \AA{}, according to
  \protect\hyperlink{Fumagalli2012}{Fumagalli et al.}
  (\protect\hyperlink{Fumagalli2012}{2012}). Our selection of sources
  in the redshift range $z=5.1$-5.4 has a mean [3.6] $-$ [4.5] color
  of 0.31 $\pm$ 0.05 mag and 0.35 $\pm$ 0.07 mag for our
  photometric-redshift and spectroscopic samples, respectively,
  implying a mean EW(H$\alpha$+[NII]+[SII]) of 665 $\pm$ 53 \AA{} and
  707 $\pm$ 74 \AA{}, respectively, for sources in these samples (and
  638 $\pm$ 118 \AA{} and 855 $\pm$ 179 \AA{}, respectively, using
  direct SED fits).  Four selected sources show even far redder colors
  than predicted with the strong evolution model. The four reddest
  sources have a mean color of 0.66 $\pm$ 0.06 mag, with a notably
  high EW of 1743 $\pm$ 221 \AA{}.  }
\label{fig:nebularlines}
\vspace{0em}
\end{figure}

The Lyman-break selection criteria applied for sources at $z$ $\sim$ 5 are
as follows:
\begin{eqnarray*}
\hspace{1.4cm}(V_{606}-i_{775}>1.2)\wedge(z_{850}-H_{160}<1.3)\wedge\\
(V_{606}-i_{775}>0.8(z_{850}-H_{160}) + 1.2)
\end{eqnarray*}
\noindent where $\wedge$ denotes the AND symbol. For non-detections,
the 1$\sigma$ upper limit is taken as the flux in the dropout
band. The aforementioned criteria enable to select sources in the
range $ z$ $\sim$ 4.5 $-$ 5.5. Therefore, a high-redshift boundary is
set by excluding sources which satisfy the selection criteria for $z$
$\sim$ 6 selection (e.g., Bouwens et al.\ 2015a).  Contamination from
sources at lower redshifts is reduced by requiring that the $z$ $\sim$
5 sources have a non-detection ($<$ 2$\sigma$) in the $B_{435}$ band.
Furthermore, we exclude point sources by requiring the SExtractor
stellarity index to be less that 0.9 (where 0 and 1 correspond to
extended and point sources, respectively). Utilizing these selection
criteria results in an initial $V$-drop sample of 1567 sources
(\hyperlink{Bouwens2015}{Bouwens et al.}
\hyperlink{Bouwens2015}{2015}).

\begin{figure}
\includegraphics[scale=1.2,trim=4mm 3mm 0mm 0mm]{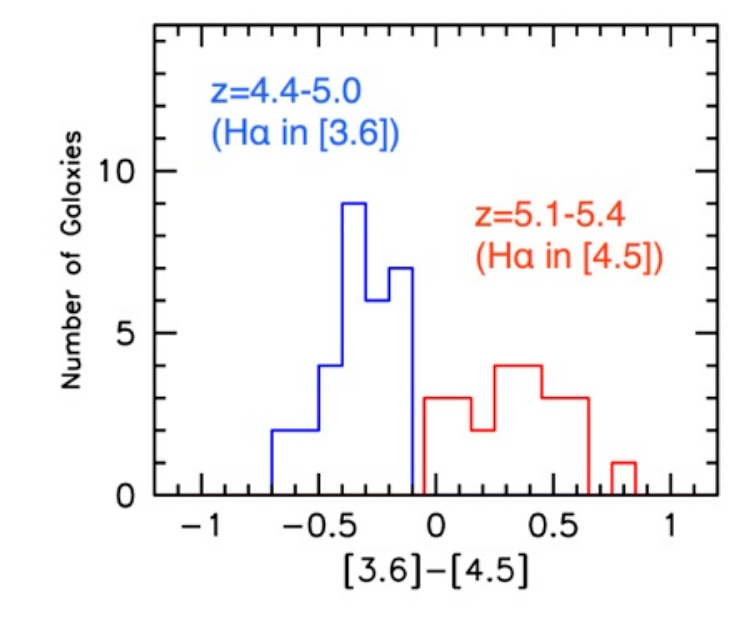}
\vspace{0em}
\caption{$[3.6]-[4.5]$ color distribution for sources in our
  $z=5.1$-5.4 primary selection (\textit{red histogram}) and
  $z=4.4$-5.0 control sample (\textit{blue histogram}: Appendix A).
  Sources where H$\alpha$+[NII] is in the [3.6] band uniformly have
  moderately blue colors, whereas sources with H$\alpha$+[NII] in the
  [4.5] band uniformly have moderately red colors.  The impact of the
  H$\alpha$+[NII] emission lines on the IRAC colors is quite clear.
  Deriving the H$\alpha$+[NII]+[SII] EW by contrasting the observed
  colors of sources in the two samples should produce a very robust
  result.}
\label{fig:col36}
\vspace{0em}
\end{figure}

In order to be able to measure the H$\alpha$ EW for $z\gtrsim5$
galaxies, it is easiest to only make use of galaxies in the narrow
redshift window $z=5.10$-5.40.  We therefore use photometric redshifts
to identify a sub-sample of galaxies in this window.  The photometric
redshifts for our sample are determined using the EAZY photometric
redshift code (\hyperlink{Brammer2008}{Brammer et al.}
\hyperlink{Brammer2008}{2008}), which compares photometric data with
synthetic photometry of galaxies for various template spectra and
redshift ranges. The best-fit redshift is then derived from a
statistical analysis of the differences between both data sets. The
aforementioned 8 HST bands ($B_{435}$, $V_{606}$, $i_{775}$,
$I_{814}$, $z_{850}$, $Y_{105}$, $J_{125}$, $H_{160}$) are used to
derive the best-fit photmetric redshifts. The IRAC photometry is
excluded from this fitting to avoid introducing any bias in the
measured [3.6] $-$ [4.5] color. This reduces the sample size to 393
sources.

Figure \ref{SED} shows an example output of EAZY for a source at $z$
$\sim$ 5.2 and a comparison of our estimated EAZY photometric
redshifts for a sample of $z$ = 4 $-$ 6 spectroscopically confirmed
sources from Stark et al. (2010: see D. Stark et al. 2015, in prep)
and Vanzella et al. (2009). The scatter around the one-to-one relation
is $\Delta z/(1+z)$ = 0.036, which provides confidence that we can
select sources in the narrow redshift range $z$ = 5.1-5.4.

We select very bright sources with the requirements that
S/N($H_{160}$) $>$ 7 $\wedge$ S($H_{160}$)/N(3.6 $\mu$m)
$>$ 3 $\wedge$ S($H_{160}$)/N(4.5 $\mu$m) $>$ 3,
where S/N is the signal to noise ratio.  Our selection of sources
based on their measured flux in the $H_{160}$ band and measured noise
in the Spitzer/IRAC bands allows us to include sources in our analysis
which we would expect to show up prominently in the Spitzer/IRAC
bands.  Basing the selection on the measured flux in the Spitzer/IRAC
  bands would bias our measurement of the colors.  We also discard 41
  sources for which the contamination by nearby objects is higher than
  200 nJy or sources with a poor IRAC deblending ($\chi^{2}$ parameter
  less than 0.2), reducing the sample to 101 bright sources.  

Finally, we require that the EAZY redshift probability distribution
prefers a redshift in the range $z=5.1$-5.4 at $>85$\%, providing a
sample with 11 sources in the redshift range $z$ $\sim$ 5.1-5.4
presented in Table \ref{table:1}.

\subsection{Spectroscopic Redshift Selection}
\label{speczsample}

In addition to making use of sources very likely to lie in the
redshift range $z$ = 5.1-5.4 using our photometry for the sources and
the redshift likelihood distributions we derive, we can also make use
of sources known to lie in the redshift range $z$ = 5.1-5.4 from
available spectroscopy (D. Stark et al. 2015, in prep).  Using
spectroscopic redshifts, we can be even more certain that the sources
we are using lie in the narrow redshift range $z$ = 5.10-5.40 required
for our desired measurement of the H$\alpha$ flux.

One potential drawback to the inclusion of such sources in the present
study is that we might be working with a biased sample, given that
essentially all of the spectroscopic redshift measurements we utilize
come from Ly$\alpha$, and it is not clear a priori that the study of
such a sample might bias the mean H$\alpha$ EW we measure to higher
values.  Fortunately, as we show in a separate study (Smit et
al. 2015, in preparation), the mean H$\alpha$+[NII]+[SII] EW measured for both
photometric-redshift and spectroscopic samples are essentially
identical.

Cross-correlating the source catalogs of
\hyperlink{Bouwens2015}{Bouwens et al.}
(\hyperlink{Bouwens2015}{2015}) and \hyperlink{Skelton2014}{Skelton et
  al.}  (\hyperlink{Skelton2014}{2014}) with the spectroscopic catalog
of D. Stark et al. (2015, in prep) and
\hyperlink{Vanzella2009}{Vanzella et al.}
(\hyperlink{Vanzella2009}{2009}), we identified 13 $z$ = 5.10 $-$ 5.40
galaxies that we can use for our study, with one source overlapping
with our photometric selection (see Table \ref{table:2}).  As in our
photometric redshift selection (\S2.2), we exclude sources where flux
from neighboring sources significantly contaminate the photometric
apertures for our $z=5.10$-5.40 sample (3 sources).  10 of the sources
in the desired redshift range were from D. Stark et al. (2015, in
prep) redshift compilation, while 3 came from the Vanzella et
al. (2009) compilation.

We will make use of sources from both our high-quality photometric
redshift sample and spectroscopic sample for the analyses that follow.

\begin{table*}
\centering
\vspace{3em}
\caption{Our selections of sources with a high probability of lying in the redshift range 5.1 $<z<$ 5.4 from the photometry.$^{\textup{a}}$}
\begin{tabular}{l|c|c|c|c|c|c|c|c} 
\hline
\hline
\newline
  &  &  &  & [3.6] $-$ [4.5] &H$\alpha$+[NII]+[SII]  & sSFR  & log$_{10}$$M$  &  \\ 
ID & RA & DEC & $z_{phot}$ & (mag) & EW [$\AA$]$^{c}$ & [Gyr$^{-1}$] & [$M$$_{\odot}$] & $M^{\textup{b}}_{UV}$ \\ 

\hline
\hline
GNDV-7133823953 &12:37:13.38 &\mbox{} 62:12:39.5 &5.2 $\pm$ 0.1 &\mbox{} 0.5 $\pm$ 0.3 &672 $\pm$ 505&30$_{-9}^{+10}$&8.84$_{-0.11}^{+0.12}$ &$-$21.3 $\pm$ 0.1 \\
GNDV-7128013231 &12:37:12.80 &\mbox{} 62:11:32.3 &5.3 $\pm$ 0.2 &\mbox{} 0.1 $\pm$ 0.2& 190 $\pm$ 254 &17$_{-6}^{+12}$&9.24$_{-0.13}^{+0.30}$ &$-$20.8 $\pm$ 0.1 \\
GNDV-7033233179$^{\textup{d}}$ &12:37:03.32 &\mbox{} 62:13:31.8 & 5.3 $\pm$ 0.1 &\mbox{} 0.4 $\pm$ 0.2&719 $\pm$ 368&24$_{-11}^{+8}$&8.93$_{-0.17}^{+0.10}$&$-$20.8 $\pm$ 0.1 \\
GNDV-6302234526 &12:36:30.22 &\mbox{} 62:13:45.3 &5.2 $\pm$ 0.1  &\mbox{} 0.3 $\pm$ 0.1&385 $\pm$ 170 &11$_{-8}^{+6}$&9.38$_{-0.28}^{+0.19}$ &$-$20.6 $\pm$ 0.1 \\
GNDV-6285841077 &12:36:28.58 &\mbox{} 62:14:10.8 &5.3 $\pm$ 0.1   &\mbox{} 0.2 $\pm$ 0.2&557 $\pm$ 449&21$_{-11}^{+17}$&8.76$_{-0.20}^{+0.32}$ &$-$20.7 $\pm$ 0.1 \\
GNWV-6514085687 &12:36:51.40 &\mbox{} 62:08:56.9 &5.3 $\pm$ 0.1 &\mbox{} 0.0 $\pm$ 0.1&25 $\pm$ 167&1$_{-1}^{+1}$&10.12$_{-0.00}^{+0.06}$ &$-$20.6 $\pm$ 0.1 \\
GNWV-6121502518 &12:36:12.15 &\mbox{} 62:10:25.2 &5.3 $\pm$ 0.1&\mbox{} 0.4 $\pm$ 0.3&917 $\pm$ 573 &6$_{-10}^{+5}$&9.29$_{-0.62}^{+0.25}$ &$-$20.8 $\pm$ 0.1 \\
GNWV-6095211615 &12:36:09.52 &\mbox{} 62:11:16.2 &5.2 $\pm$ 0.1 &\mbox{} 0.4 $\pm$ 0.1&913 $\pm$ 274 &19$_{-10}^{+12}$&9.24$_{-0.20}^{+0.27}$ &$-$21.6 $\pm$ 0.1 \\
GNDV-3756634257 &12:37:05.66 &\mbox{} 62:13:42.6 &5.3 $\pm$ 0.1 &\mbox{} 0.1 $\pm$ 0.1&274 $\pm$ 180 &26$_{-19}^{+11}$&9.55$_{-0.29}^{+0.15}$ &$-$21.2 $\pm$ 0.1 \\
GNDV-6325033158 &12:36:32.50 &\mbox{} 62:13:31.6 &5.3 $\pm$ 0.1 &\mbox{} 0.4 $\pm$ 0.1&882 $\pm$ 262 &9$_{-3}^{+5}$&8.79$_{-0.07}^{+0.24}$ &$-$20.9 $\pm$ 0.1 \\
GSDV-2332672480  &03:32:33.26 &$-$27:47:24.8      &5.3   $\pm$ 0.1  &\mbox{} 0.6 $\pm$ 0.1&1480 $\pm$ 110&1$_{-1}^{+1}$&10.06$_{-0.01}^{+0.06}$ &$-$20.6 $\pm$ 0.1 \\
\hline
\multicolumn{8}{p{.8\textwidth}}
{\small $^{\textup{a}}$ To identify those sources with the highest probability of lying in the redshift range $z$ = 5.10 $-$ 5.40, sources are required to have $P(5.1<z<5.4)>0.85$.  \newline 
\small $^{\textup{b}}$ The $z_{850}$ band magnitude is used to derive the intrinsic UV luminosity. \newline 
\small $^{\textup{c}}$ The estimated EW for individual sources is derived by comparing the 4.5 $\micron$ flux with that derived from FAST (excluding the 4.5 $\micron$ flux from the fits).\newline 
\small $^{\textup{d}}$ Spectroscopically confirmed to be at $z$ = 5.21 (D. Stark et al. 2015, in prep). \newline
}
\end{tabular}
\label{table:1}
\end{table*}

\begin{table*}
\centering
\caption{Sample of spectroscopically confirmed sources in the redshift range 5.1 $< z <$ 5.4.$^{\textup{a}}$}
\begin{tabular}{l c c c c  c  c c c c}
\hline
\hline
\newline
\mbox{}
\newline
  &   &   &   & &  [3.6] $-$ [4.5] & H$\alpha$+[NII]+[SII]  & sSFR  & log$_{10}$$M$  &  \\
ID & RA & DEC & $z_{spec}^{\textup{a}}$ & $z_{phot}$ & (mag) & EW [$\AA$]$^{b}$ & [Gyr$^{-1}$] &  [$M$$_{\odot}$] & $M_{UV}$ \\
\hline
\hline
GNDV-6554953313 &12:36:55.49 &62:15:33.1    & 5.19 &4.9 $\pm$ 0.1	 &0.6   $\pm$  0.2	&1579  $\pm$ 359 &10$_{-7}^{+6}$ &9.09$_{-0.28}^{+0.24}$ &$-$21.3 $\pm$ 0.1\\
GNDV-7033233179 &12:37:03.32 &62:13:31.8    &5.21  &5.3 $\pm$ 0.1 &0.4     $\pm$  0.2		&719  $\pm$ 368 &24$_{-11}^{+8}$ &8.93$_{-0.17}^{+0.10}$ &$-$20.8 $\pm$ 0.1\\
GNDV-7027322916 &12:37:02.73 &62:12:29.2     &5.23	&5.5 $\pm$ 0.1 & 0.5    $\pm$  0.1	&1000  $\pm$ 257 &37$_{-16}^{+12}$ &9.21$_{-0.13}^{+0.06}$ &$-$20.6 $\pm$ 0.1\\
GNDV-6375223629 &12:36:37.52 &62:12:36.3     &5.18	 &5.4 $\pm$ 0.1 &0.0  $\pm$   0.3	& $<$419$^{\textup{c}}$ &21$_{-35}^{+19}$ &8.62$_{-0.71}^{+0.35}$ &$-$20.0 $\pm$ 0.1\\
GNDV-6553954912 &12:36:55.39 &62:15:49.1     &5.19	 &5.1 $\pm$ 0.2 &0.8    $\pm$  0.4	&2568 $\pm$ 1561 &7$_{-5}^{+9}$ &8.29$_{-0.23}^{+0.52}$ &$-$19.8 $\pm$ 0.1\\
GNWV-7347782930 &12:37:34.77 &62:18:29.3     &5.32 &5.2 $\pm$ 0.2  &0.6    $\pm$  0.3	&1344  $\pm$ 666 &3$_{-3}^{+4}$ &8.90$_{-0.42}^{+0.44}$ &$-$19.9 $\pm$ 0.1\\
ERSV-2213040511 & 03:32:21.30 & $-$27:40:51.2&5.29 &5.3 $\pm$ 0.1 &0.5 $\pm$ 0.2 &1075 $\pm$ 394 &18$_{-10}^{+13}$ &9.11$_{-0.18}^{+0.27}$ &$-$20.9 $\pm$ 0.1\\
GSWV-2454254386 & 03:32:45.43 & $-$27:54:38.6&5.38 &5.4 $\pm$ 0.1 &0.2 $\pm$ 0.1 &289 $\pm$ 75 &54$_{-82}^{+69}$ &8.96$_{-0.66}^{+0.55}$ &$-$21.6 $\pm$ 0.1\\
GND6418$^{\textup{d}}$ & 12:36:18.19   & 62:10:21.9   &5.28 &5.6 $\pm$ 0.1  &0.3   $\pm$  0.1 &309  $\pm$ 88 &24$_{-8}^{+14}$  &9.99$_{-0.01}^{+0.20}$ &$-$21.6 $\pm$ 0.1\\
GND33928 & 12:37:36.87   & 62:18:55.9   &5.35 &5.2 $\pm$ 0.1 &0.1  $\pm$  0.2 & $<$222$^{\textup{c}}$ &19$_{-16}^{+17}$ &9.09$_{-0.19}^{+0.23}$ &$-$20.4 $\pm$ 0.1\\
GND29175 & 12:37:31.45   & 62:17:08.3   &5.25 &5.3 $\pm$ 0.2 &0.3   $\pm$  0.4 &666  $\pm$ 929 &3$_{-7}^{+4}$  &9.28$_{-0.84}^{+0.31}$ &$-$19.9 $\pm$ 0.1\\
GND12038 & 12:36:26.49  & 62:12:07.4	&5.20 &5.2 $\pm$ 0.4 &0.0   $\pm$  0.2 &254  $\pm$ 384 &10$_{-9}^{+12}$ &9.11$_{-0.26}^{+0.43}$ &$-$20.3 $\pm$ 0.1\\
GS48361$^{\textup{d}}$ & 03:32:16.55 & $-$27:41:03.2&5.25 &5.5 $\pm$ 0.2 &0.3 $\pm$ 0.2 &668 $\pm$ 370 &6$_{-9}^{+6}$ &9.32$_{-0.53}^{+0.10}$ &$-$20.7 $\pm$ 0.1\\
\hline
\multicolumn{8}{p{.8\textwidth}}{\small $^{\textup{a}}$ The spectroscopic redshifts are obtained by cross-correlating the \hyperlink{Bouwens2015}{Bouwens et al.} (\hyperlink{Bouwens2015}{2015}) and \hyperlink{Skelton2014}{Skelton et al.} (\hyperlink{Skelton2014}{2014}) catalogs with the spectroscopic catalog of D. Stark et al. (2015, in prep) and Vanzella et al.\ (2009).\newline
\small $^{\textup{b}}$ The estimated EW for individual sources is derived by comparing the 4.5 $\micron$ flux with that derived from FAST (excluding the 4.5 $\micron$ flux from the fits).\newline 
\small $^{\textup{c}}$ We give the error in the H$\alpha$+[NII]+[SII] EW as a 1$\sigma$ upper limit when the inferred value is negative. \newline
\small $^{\textup{d}}$ Spectroscopic redshift measurement is based on the identification of a probable absorption line and hence less confident than the other spectroscopic redshift measurements included in this table (D. Stark et al.\ 2015, in prep; Vanzella et al.\ 2009).\newline
}
\end{tabular}\\
\label{table:2}
\end{table*}

\section{Results}

Using the selection discussed in \S2.2 and \S2.3 we have isolated a sample of 21 galaxies that have redshifts in the narrow redshift range $z$ $\sim$ 5.1 $-$ 5.4 at high confidence. In this section we will discuss the IRAC [3.6] $-$ [4.5] colors of these galaxies and compare them with the IRAC colors of a spectroscopically confirmed sample of $z$ $\sim$ 4.5 galaxies.

\subsection{Mean [3.6] $-$ [4.5] Color for $z=5.1-5.4$ Galaxies}

Our selection of galaxies in the redshift range $z$ $\sim$ 5.1 $-$ 5.4 allows us to solve for the rest-frame EW of the nebular emission lines using the [3.6] $-$ [4.5] color. Assuming $F_{\nu}$ $\approx$ constant  (i.e. $F_{\lambda}$ $\propto$ $\lambda^{-2}$) we approximate the observed flux $F_{\nu, \textup{obs}}$ in the IRAC filters, for a given EW, by 

\begin{gather}
F_{\nu , \textup{obs}} = F_{\nu , \textup{continuum}}\cdot x \notag \\
 x_{\rm EW}=\left(1+\sum_{{\rm lines},\textit{i}} \frac{\rm{EW_{0,\textit{i}}}\cdot(1+z)\cdot R(\lambda_{{\rm obs},\textit{i}})}{\lambda_{\rm obs,\textit{i}}\int R(\lambda)/\lambda\, d\lambda}\right),
\end{gather}

where $F_{\nu , \textup{continuum}}$ is the stellar continuum flux and $\lambda_{\textup{obs},i}$ is the observed wavelength of the nebular emission lines (\hyperlink{Smit2014}{Smit et al.}\hyperlink{Smit2014}{2014}). $R(\lambda)$ denotes the response curve of the filter. 
 We use all nebular emission lines tabulated in \hyperlink{AndersFA2003}{Anders \& Fritze-v. Alvensleben} (\hyperlink{AndersFA2003}{2003}) and the hydrogen Balmer lines for the modeling of the [3.6]$-$ [4.5] color. We fix the line intensities relative to H$\beta$ according to the values in \hyperlink{AndersFA2003}{Anders \& Fritze-v. Alvensleben} (\hyperlink{AndersFA2003}{2003}) for sub-solar metallicity 0.2 Z$_{\odot}$ and assuming case B recombination. The observed [3.6] $-$ [4.5] color can then be modeled by

\begin{equation}
[3.6]-[4.5] = ([3.6]-[4.5])_{continuum} - 2.5 \log_{10} \left ( \frac{x_{3.6}}{x_{4.5}} \right ).
\end{equation}

Figure~\ref{fig:nebularlines} illustrates the contamination by nebular
emission lines in the photometric filters as a function of
redshift. The top panel shows the redshift ranges where the dominant
emission lines, H$\alpha$, H$\beta$, [OIII]$\lambda\lambda$4959,
5007, [NII], and [SII] fall in the IRAC 3.6 $\mu$m and 4.5 $\mu$m passbands. The lower
panel shows the expected [3.6] $-$ [4.5] color in the presence of
nebular line emission. The dotted line illustrates the case when the
rest frame EW evolves strongly with redshift. Consistent with
\hyperlink{Fumagalli2012}{Fumagalli et al.}
(\hyperlink{Fumagalli2012}{2012}) who find that EW$_{0}$(H$\alpha$)
$\propto$ (1+$z$)$^{1.8}$ \AA{} for star forming galaxies at 0
$\lesssim$ $z$ $\lesssim$ 2.  Galaxies are expected to become quite
red in the redshift range $z$ = 5.1 $-$ 5.4, due to contamination of
the 4.5 $\mu$m flux by the H$\alpha$ line and no strong nebular emission
line contamination in the 3.6 $\mu$m band.

The selected galaxies in the redshift range $z$ $\sim$ 5.1 $-$ 5.4
allow us to deduce the EW of the nebular emission lines in $z$
$\gtrsim$ 5 objects from the fluxes in the 3.6 $\mu$m and 4.5 $\mu$m
bands.  We observe a mean [3.6] $-$ [4.5] color of 0.31 $\pm$ 0.05 mag
for the selected sources in our photometric-redshift sample and a mean
[3.6] $-$ [4.5] color of 0.35 $\pm$ 0.07 mag for sources in our
spectroscopic sample.  In both cases, we estimate the uncertainty here
through bootstrap resampling.  

We estimate the [3.6] $-$ [4.5] continuum color by contrasting the
mean [3.6] $-$ [4.5] colors observed for $z=4.4$-5.0
spectroscopically-confirmed sources from
\hyperlink{Vanzella2009}{Vanzella et al.}
(\hyperlink{Vanzella2009}{2009}), \hyperlink{Shim2011}{Shim et al.}
(\hyperlink{Shim2011}{2011}), and \protect\hyperlink{Stark2013}{Stark
  et al.}  (\protect\hyperlink{Stark2013}{2013}), i.e., $-0.32\pm0.03$
mag, with the mean [3.6]$-$[4.5] colors observed for our selected $z$
$\sim$ 5.1 $-$ 5.4 sample and attribute any differences in the
[3.6]$-$ [4.5] colors to the impact of flux from the H$\alpha$ + [NII]
+ [SII] lines.  We estimate a color for the stellar continuum of [3.6]
$-$ [4.5] of $0.00\pm0.04$ mag.  See Figure~\ref{fig:col36} for an
illustration of the differences between the two samples.

Four sources in our selection show considerably redder [3.6] $-$ [4.5]
colors than we can infer from our model SED with strong line emission,
i.e., 0.66 $\pm$ 0.06 mag.  The very red [3.6] $-$ [4.5] colors
observed for many sources in our $z$ = 5.1 $-$ 5.4 sample suggests
that the IRAC [3.6] $-$ [4.5] color may provide significant leverage
in terms of identifying galaxies that specifically lie in the narrow
redshift range $z$ $\sim$ 5.1 $-$ 5.4. This method has previously been
explored at $z\sim7-8$ by \hyperlink{Smit2015}{Smit et al.}
(\hyperlink{Smit2015}{2015}) and \hyperlink{RB2015}{Roberts-Borsani et
al.} (\hyperlink{RB2015}{2015}) based on the impact of [OIII] on the IRAC
fluxes at these high redshifts.

\begin{figure*}
\centering
\includegraphics[scale=0.8,trim=14mm 5mm 0mm 0mm]{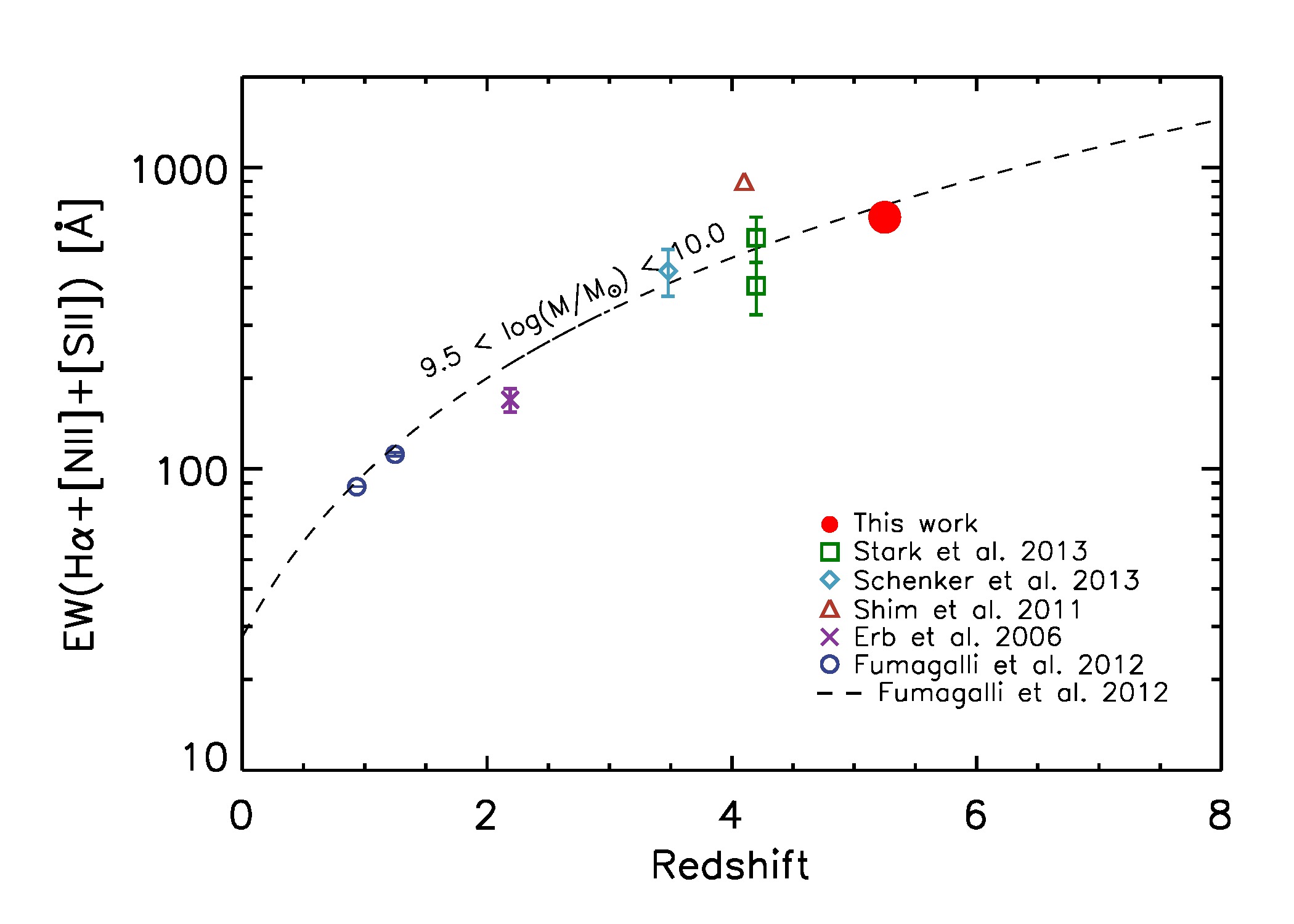}
\caption{\label{EW}H$\alpha$+[NII]+[SII] EWs as function of redshift
  derived from the mean [3.6] $-$ [4.5] color of our source
  selection. Several estimates from the literature are indicated for
  reference 
  (\protect\hyperlink{Erb2006}{Erb et al.}
  \protect\hyperlink{Erb2006}{2006}; \protect\hyperlink{Shim2011}{Shim
    et al.} \protect\hyperlink{Shim2011}{2011};
  \protect\hyperlink{Fumagalli2012}{Fumagalli et al.}
  \protect\hyperlink{Fumagalli2012}{2012};  \protect\hyperlink{Schenker2013}{Schenker
  et al.}  \protect\hyperlink{Schenker2013}{2013};\protect\hyperlink{Stark2013}{Stark et
    al.} \protect\hyperlink{Stark2013}{2013}).  The upper and lower EW
  determination for Stark et al. (2013) excludes and does not exclude
  the contaminated IRAC band in deriving the stellar continuum
  required to derive the EW for the H$\alpha$+[NII]+[SII] line.  
  We assume line ratios as listed by
  \protect\hyperlink{AndersFA2003}{Anders \& Fritze-v. Alvensleben}
  (\protect\hyperlink{AndersFA2003}{2003}) and that the stellar
  continuum of our $z=5.10$-5.40 sample has a [3.6] $-$ [4.5] color of
  0.00 $\pm$0.04 magnitude. Redder [3.6] $-$ [4.5] colors will
  therefore be due to the emission lines contaminating the 4.5 $\mu$m
  flux.  The measured H$\alpha$+[NII]+[SII] EW shown here represents
  the weighted mean of our EW estimates from our photometric and
  spectroscopic $z=5.1$-5.4 samples (i.e., 684$\pm$51 \AA) and is
  higher than values derived at lower redshifts, suggesting stronger
  line emission at $z$ $\sim$ 5. The evolution of
  H$\alpha$+[NII]+[SII] EW for the indicated stellar mass range found
  by \protect\hyperlink{Fumagalli2012}{Fumagalli et al.}
  (\protect\hyperlink{Fumagalli2012}{2012}) is extrapolated (given by
  the dashed line) and is consistent with our inferred
  H$\alpha$+[NII]+[SII] EW. }
\label{fig:EW}
\end{figure*}

\subsection{Mean H$\alpha$+[NII]+[SII] EW of $z\sim5$ Galaxies}

By comparing the mean [3.6] $-$ [4.5] color of our $z=5.10$-5.40 sample
with the mean [3.6] $-$ [4.5] color of our $z=4.4$-5.0 sample, we find
an overall color difference of 0.68$\pm$0.08 mag relative to our
$z=5.1$-5.4 photometric sample and 0.69$\pm$0.09 mag relative to our
$z=5.1$-5.4 spectroscopic sample.  Comparing the color difference we
observe with that predicted based on simple model spectra with an
H$\alpha$ EW of 595 \AA$\,$ and line ratios set by the
\hyperlink{AndersFA2003}{Anders \& Fritze-v. Alvensleben}
(\hyperlink{AndersFA2003}{2003}) model (\textit{dotted line} in Figure
2), we infer an approximate H$\alpha$ EW of 557 $\pm$ 44 $\AA$ for our
$z=5.1$-5.4 photometric sample and 592 $\pm$ 62 $\AA$ for our
$z=5.1$-5.4 spectroscopic sample.  

Given that flux in the [NII] and [SII] lines would add to the color
difference observed between our $z=4.4$-5.0 and $z=5.1$-5.4 samples
and cannot be determined separately, it is best to quote a constraint
on the mean EW of H$\alpha$+[NII]+[SII].
\hyperlink{AndersFA2003}{Anders \& Fritze-v. Alvensleben}
(\hyperlink{AndersFA2003}{2003}) predict a contribution of 6.8\% from
          [NII] and 9.5\% from [SII].  This is in good agreement with
          an observed ratio of [NII]/H$\alpha$ of 0.05-0.09 in
          $z\sim2.3$ galaxies with stellar masses in the range
          $\log(M_\ast/M_\odot)=9.15-9.94$ by
          \hyperlink{Sanders2015}{Sanders et
            al.}(\hyperlink{Sanders2015}{2015}).  If we correct for
          this, the mean EW in H$\alpha$+[NII]+[SII] is 665 $\pm$ 53
          \AA{} for our photometric sample and 707 $\pm$ 74 \AA{} for
          our spectroscopic sample.

We can also derive H$\alpha$+[NII]+[SII] EWs for individual sources in
our photometric and spectroscopic $z=5.1$-5.4 samples.  In computing
the EWs for individual sources, we use FAST to fit the observed SEDs
of individual sources excluding the 4.5 $\micron$ band which is
contaminated by H$\alpha$ emission.  Then, by comparing the observed
4.5 $\micron$ flux with the expected 4.5 $\micron$ flux (without
including emission lines in the FAST modeling), we derive EWs for
individual sources.  The results are presented in Tables 1 and 2.  The
mean H$\alpha$+[NII]+[SII] EW we derive for our photometric sample is
638 $\pm$ 118 \AA{}, while we find 855 $\pm$ 179 \AA{} for our
spectroscopic sample.  If we follow the model results from
\hyperlink{AndersFA2003}{Anders \& Fritze-v. Alvensleben}
(\hyperlink{AndersFA2003}{2003}) and suppose that 16.3\% of the 4.5
$\micron$ excess derives from [NII] and [SII], the excesses we derive
suggest H$\alpha$ EWs of 534 $\pm$ 99 \AA{} and 715 $\pm$ 150 \AA{},
respectively.

The four reddest sources have a mean [3.6] $-$ [4.5] color of 0.66
$\pm$ 0.06 mag, consistent with an EW$_{0}$(H$\alpha$+[NII]+[SII]) of
1743 $\pm$ 221 \AA{}, equivalent to EW$_{0}$(H$\alpha$) $=$ 1458 $\pm$
185 \AA{} for the above line ratios.

Figure~\ref{fig:EW} shows several values for the H$\alpha$ + [NII] +
[SII] EWs with redshift from the literature. The black line gives the
evolution of the H$\alpha$ + [NII] + [SII] derived by
\hyperlink{Fumagalli2012}{Fugamalli et al.}
(\hyperlink{Fumagalli2012}{2012}) 
for galaxies with masses $M$ $\sim$
10$^{10}$ $-$ 10$^{11.5}$ $M_{\odot}$, which we extrapolate to higher
redshifts and lower masses. Keeping in mind that the EW(H$\alpha$ + [NII] + [SII]) as
function of the redshift is higher for sources with lower steller
masses 
our result is consistent with the
extrapolation from \hyperlink{Fumagalli2012}{Fumagalli et al.}
(\hyperlink{Fumagalli2012}{2012}) and the high EWs derived by
\hyperlink{Shim2011}{Shim et al.} (\hyperlink{Shim2011}{2011}),
\hyperlink{Stark2013}{Stark et al.} (\hyperlink{Stark2013}{2013}) and \hyperlink{Schenker2013}{Schenker
  et al.}  (\hyperlink{Schenker2013}{2013}).

Uncertainties in the photometric redshifts for our $z=5.10$-5.40
sample can lead to a systematic underestimate of the
H$\alpha$+[NII]+[SII] flux, if it causes us to include sources which
lie outside the desired range.  For $z<5.1$ sources, the 3.6 $\mu$m
band will be contaminated by H$\alpha$ + [NII] + [SII] emission.
Meanwhile, for $z>5.4$ sources, H$\beta$ + [OIII] emission will
contribute to the $3.6\mu$m band.  In both cases, the [3.6] $-$ [4.5]
color will be much bluer, causing us to infer a substantially lower EW
for that source, than is truly present.

\begin{figure*}
\centering
\includegraphics[scale=0.75,trim=14mm 5mm 0mm 0mm]{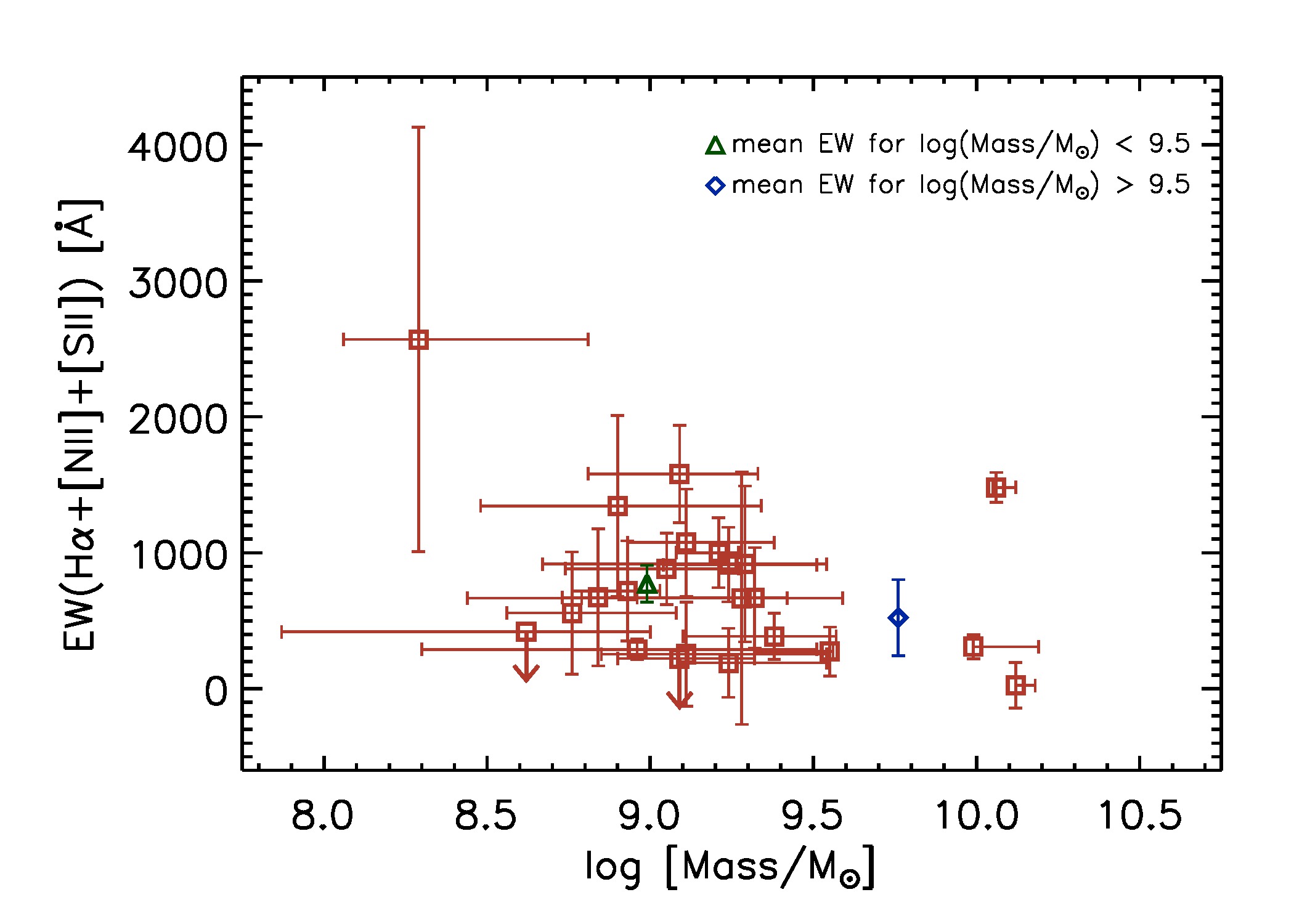}
\caption{The estimated H$\alpha$+[NII]+[SII] equivalent width from the
  4.5 $\mu$m excess vs. the estimated stellar mass (red squares with
  $1\sigma$ error bars).  The mean H$\alpha$+[NII]+[SII] EW we
  estimate for sources with an estimated stellar mass $>$
  10$^{9.5}$$M$$_{\odot}$ is 522 $\pm$ 279 \AA{} (blue diamond), while
  the mean H$\alpha$+[NII]+[SII] EW we estimate for sources with an
  estimated stellar mass $<$ 10$^{9.5}$$M$$_{\odot}$ is 773 $\pm$ 136
  \AA{} (green triangle).  The present results provide no strong
  evidence for a correlation between the inferred
  H$\alpha$+[NII]+[SII] EWs and the stellar mass. However, we
  emphasize that this may change as the samples and dynamical masses
  become larger.}
\label{fig:EWmass}
\end{figure*}

\subsection{Specific Star Formation Rates}

A clean measurement of the stellar continuum emission is essential for
deriving the sSFR. In our target redshift range the flux in the 3.6
$\mu$m band can be used for this purpose, while the 4.5 $\mu$m band is
contaminated by emission lines and should be left out in stellar
population modeling. We derive the mean sSFR for our selected sources
using similar method as described in \hyperlink{Stark2013}{Stark et
  al.}  (\hyperlink{Stark2013}{2013}).  We derive stellar masses using
the modeling code FAST (\hyperlink{Kriek2009}{Kriek et al.}
\hyperlink{Kriek2009}{2009}), which fits stellar population synthesis
templates from \hyperlink{BruzualCharlot2003}{Bruzual \& Charlot}
(\hyperlink{BruzualCharlot2003}{2003}) to broadband photometry. For
consistency, we assume a \hyperlink{Salpeter1955}{Salpeter}
(\hyperlink{Salpeter1955}{1955}) IMF with 0.1$-$100 $M_{\odot}$, a
sub-solar metallicity $Z$ = 0.2 $Z_{\odot}$ and dust attenuation from
\hyperlink{Calzetti2000}{Calzetti et al.}
(\hyperlink{Calzetti2000}{2000}). The ages range from 10 Myr to the
age of the universe at the redshift of the source. The star formation
history is assumed to be constant and the dust content is varied
between A$_{V}$ = 0 $-$ 3 mag. In order to avoid the degeneracy
between the ages of the galaxies and their dust content in the FAST
modeling of the sSFR, we derive star formation rates directly from the
UV continuum, using the \hyperlink{Kennicutt1998}{Kennicutt}
(\hyperlink{Kennicutt1998}{1998}) relation and fixing the dust
extinction using the relationship from \hyperlink{Meurer1999}{Meurer
  et al.} (\hyperlink{Meurer1999}{1999}) between A$_{V}$ and the
UV-continuum slope $\beta$.  From this method, we derive a median sSFR
of $17_{-5}^{+2}$ Gyr$^{-1}$ (individual sSFR estimates are listed in
Table \ref{table:1} and Table \ref{table:2}).

We also explore the use of a SMC dust-law, which might seem more
appropriate for Lyman-break galaxies at $z\sim5-6$ based on recent
ALMA results by \hyperlink{Capak2015}{Capak et al.}
(\hyperlink{Capak2015}{2015}). We fit FAST models as above with a
\hyperlink{Noll2009}{Noll et al.}  (\hyperlink{Noll2009}{2009}) dust
law with parameters $E_b=0.01$ and $\delta=-0.42$ and find a sSFR of
$\sim13$ Gyr$^{-1}$.

It is useful to compare the median sSFR estimates we find here,
$17_{-5}^{+2}$ Gyr$^{-1}$ and 13 Gyr$^{-1}$, with previous estimates.
This represents a $7_{-2}^{+1}$ $\times$ increase in sSFR compared to
the 2.4 Gyr$^{-1}$ value found at $z\sim2$
(\hyperlink{Reddy2012}{Reddy et al.}  \hyperlink{Reddy2012}{2012a}),
supporting a significant evolution in the sSFR.  The typical stellar
mass for galaxies in our $z=5.1$-5.4 selection is $\sim$10$^9$
$M_{\odot}$, so we will make our comparison with previous measures at
this stellar mass.  Both Stark et al.\ (2013) and Gonzalez et
al.\ (2014) find that the typical galaxy with this stellar mass has a
$UV$ luminosity $M_{UV}$ of $-$20 mag.  Accounting for a factor of
$\sim$2 mean dust attenuation at this luminosity (as implied by the
Bouwens et al.\ 2014 $\beta\sim-1.9$ results), the equivalent sSFR is
$\sim$11 Gyr$^{-1}$.  Somewhat similarly, Salmon et al.\ (2015) derive
a sSFR of $\sim$8$_{-4}^{+8}$ Gyr$^{-1}$, which is again somewhat
lower than we find here.

All things being equal, we would expect the sSFR estimates we derive
here to be more accurate than previous estimates, given our precise
knowledge of the redshifts and hence the position of nebular emission
lines within galaxies.  The impact of the lines on the mass and sSFR
estimates could be as large as a factor of $\sim$1.5, allowing for an
approximate reconciliation of the present sSFR estimates with that from
previous work.  However, the present sample of $z\sim5.1$-5.4 galaxies
is still quite small, and therefore expansion of the present sample
would certainly be helpful for improving our sSFR estimate.

\subsection{Possible Dependence of the EW of H$\alpha$+[NII]+[SII] on the Stellar Mass} 

By fitting to the photometry of all passbands uncontaminated by the
strong H$\alpha$+[NII]+[SII] nebular emission lines, we can estimate
stellar masses for sources in our $z=5.1$-5.4 samples, as described in the
previous section.  As these sources are distributed over a wide range
of stellar mass, i.e., $10^{8.5}$ $M_{\odot}$ to $10^{10.4}$
$M_{\odot}$, we can go beyond a simple determination of the mean
H$\alpha$+[NII]+[SII] EW for $z\sim5$ galaxies and look at whether
there is a dependence on stellar mass.

Any significant dependence on stellar mass would be noteworthy, as it
could point to a significant mass or scale dependence to the star
formation histories of galaxies.  While such a scale dependence could
be expected if there are significant feedback effects at early times
(e.g., Bowler et al.\ 2014; Bouwens et al.\ 2015), many simulations
(e.g., Finlator et al.\ 2011) predict that galaxies build up their
stellar mass in a relatively self-similar manner, independent on the
overall stellar mass.

We examine the evidence for such a correlation between the
H$\alpha$+[NII]+[SII] EW and the stellar mass in
Figure~\ref{fig:EWmass}, plotting the H$\alpha$+[NII]+[SII] EWs we
estimate for individual sources against the stellar masses we estimate
for the same sources.  Also shown on this figure is the average
H$\alpha$ EW we measure for galaxies with estimated stellar masses
$<$10$^{9.5}$ $M_{\odot}$ (773 $\pm$ 136 \AA{}) and for those with
estimated stellar masses $>$10$^{9.5}$ $M_{\odot}$ (522 $\pm$ 279
\AA{}).

At face value, these results do not provide any strong evidence
($<2\sigma$) for a correlation between the H$\alpha$ EW and the
stellar mass.  While it is possible that a slight correlation might be
expected (i.e., since massive galaxies would be the first to
experience a slowing in their growth rate due to feedback-type
effects), no strong trend is evident.  A similar conclusion can be
drawn by fitting H$\alpha$+[NII]+[SII] EWs vs. stellar mass relation to a
straight line.

\section{Discussion and Summary}

In this paper, we derive the H$\alpha$+[NII]+[SII] EW and the sSFR by
selecting galaxies at $z$ $\sim$ 5.1-5.4.  In doing so, we make use of
the highest redshift window allowing for a clean measurement of both
the H$\alpha$ flux and the stellar continuum.  Reliable estimates of
the stellar mass and sSFR require a clean measurement of the stellar
continuum.  At $z=5.4$-6.6, rest-frame optical flux information on
galaxies -- as derived from the Spitzer/IRAC data -- can be quite
ambiguous to interpret, due to a significant contribution from nebular
emission lines (H$\alpha$, H$\beta$, [OIII], [NII], [SII]) in both
sensitive Spitzer/IRAC channels (i.e., [3.6] and [4.5]), and at
$z>6.6$, the H$\alpha$ line redshifts out of the [4.5] channel.

Observing galaxies in the $z=5.1$-5.4 redshift interval is useful as a
contrast and for interpreting the observations of galaxies in the
redshift interval $z=3.8$-5.0.  In both intervals, the Spitzer/IRAC
fluxes are expected to be dominated by a stellar continuum
contribution and a contribution from H$\alpha$.  However, the
H$\alpha$ line will contribute to the measured flux in a different
Spitzer/IRAC band at $z=5.1$-5.4 ([4.5]) than at $z=3.8$-5.0 ([3.6]),
so the colors of galaxies in the two samples can be contrasted and
used to set strong constraints on the rest-frame EW of H$\alpha$.

In our utilization of the $z=5.1$-5.4 redshift interval to study the
H$\alpha$ EW and sSFR, we have selected 11 bright sources over the
CANDLES GOODS-North and GOODS-South Fields satisfying the criteria
S($H_{160}$)/N(3.6 $\mu$m) $>$ 3, S($H_{160}$)/N(4.5 $\mu$m) $>$ 3,
($V_{606} - i_{775}$ $>$ 1.2 ), ($z_{850} - H_{160}$ $<$ 1.3),
($V_{606} - i_{775}$ $>$ 0.8 ($z_{850} - H_{160}$) + 1.2), and
S/N($B_{435}$) $<$ 2.  The candidates are required to have $>$ 85\%
probability of lying in the redshift range $z$ $\sim$ 5.1 $-$ 5.4.  We
have supplemented this sample with 13 $z$ = 5.10 $-$ 5.40 sources from
the spectroscopic redshift sample of Vanzella et al.\ (2009) and
D. Stark et al.\ (2015, in prep).

We find a mean rest-frame EW(H$\alpha$+[NII]+[SII]) of 665 $\pm$ 53
\AA{} for our photometric sample and 707 $\pm$ 74 \AA{} for our
spectroscopic sample based on the mean [3.6] $-$ [4.5] colors of these
samples.  Assuming that 84\% of the H$\alpha$+[NII]+[SII] line flux is
H$\alpha$, we further derive a H$\alpha$ EW of $\sim$ 557$\pm$44 \AA{}
and 592 $\pm$62 \AA{} for our photometric and spectroscopic sample,
respectively.  Our estimate is consistent with the (1+$z$)$^{1.8}$
power law derived for a strong line-emitter model by
\hyperlink{Fumagalli2012}{Fumagalli et al.}
(\hyperlink{Fumagalli2012}{2012}). Four sources have an even higher EW
than predicted by the strong line-emitter model in
\hyperlink{Fumagalli2012}{Fumagalli et al.}
(\hyperlink{Fumagalli2012}{2012}).  These reddest sources in our
selection have a mean [3.6] $-$ [4.5] color of 0.66 $\pm$ 0.06 mag
corresponding to an average EW of 1743 $\pm$ 221 \AA{}.

Our selection at $z$ $\sim$ 5 has a median sSFR of $\sim$
17$_{-5}^{+2}$ Gyr$^{-1}$.  This represents a $7_{-2}^{+1}$ $\times$ increase in
sSFR compared to the 2.4 Gyr$^{-1}$ value found at $z\sim2$
(\hyperlink{Reddy2012}{Reddy et al.}  \hyperlink{Reddy2012}{2012a}),
supporting a significant evolution in the sSFR. Our estimate is in
agreement with the theoretical model of
\hyperlink{NeisteinDekel2008}{Neistein \& Dekel}
(\hyperlink{NeisteinDekel2008}{2008}), matching the increasing
specific inflow rate of baryonic particles. 

We emphasize the H$\alpha$+[NII]+[SII] EWs and sSFR we derive here for our
photometric sample is effectively a lower limit on the true value, as
the inclusion of any sources in our photometric selection outside the
desired redshift range due to photometric redshift uncertainties would
result in a bluer mean [3.6] $-$ [4.5] color and higher redshift
optical flux for the average source.

We also take advantage of the stellar population modeling we do of
individual sources in our sample and the range of estimated masses to
look at a possible correlation between the H$\alpha$+[NII]+[SII] EW of
sources in our selection and their stellar masses.  We find no strong
evidence ($<$2$\sigma$) for there being a correlation between the EW
of H$\alpha$+[NII]+[SII] EWs and the stellar mass.  However, we
caution that our sample sizes and dynamic range are limited, and so a
correlation may be evident when examining large sample sizes or a
wider dynamic range.

More accurate results would require spectroscopic redshifts for a
larger number of sources. Though our target redshift range allows for
a relatively clean measurement, constructing large samples of
$z\sim$ 5.1 $-$ 5.4 galaxies is challenging due to the narrow redshift
window we are considering. We expect significant progress in the
future as a result of future samples with the MUSE spectrograph (Bacon
et al.\ 2015). Furthermore, ultra-deep Spitzer/IRAC data over the GOODS-N 
and GOODS-S fields will become available through the GOODS Re-ionization 
Era wide-Area Treasury from Spitzer (GREATS, PI: Labb\'e) program (2014).

\section*{Acknowledgments}
We acknowledge the support of NASA grant NAG5-7697, NASA/STScI grant
HST-GO-11563, and a NWO vrij-competitie grant 600.065.140.11N211.  RS
acknowledges the support of the Leverhulme Trust.  This research has
made use of the NASA/IPAC Infrared Science Archive, which is operated
by the Jet Propulsion Laboratory, California Institute of Technology,
un- der contract with the National Aeronautics and Space
Administration.

\appendix

\section{$z = 4.4 - 5.0$ Reference Sample}

In order to estimate the mean stellar continuum color $[3.6]-[4.5]$
for $z\sim5$ (\S3.1), we make use of a spectroscopic selection of
sources at $z=4.4$-5.0 from Vanzella et al.\ (2009), Shim et
al.\ (2011), and Stark et al. (2013). We obtain a sample of 30 $z =
4.4$-5.0 sources by cross-correlating the source catalogs of
\hyperlink{Bouwens2015}{Bouwens et al.}
(\hyperlink{Bouwens2015}{2015}) and \hyperlink{Skelton2014}{Skelton et
  al.}  (\hyperlink{Skelton2014}{2014}) with the spectroscopic catalog
of Shim et al. (2011), Stark et al. (2013) and
\hyperlink{Vanzella2009}{Vanzella et al.}
(\hyperlink{Vanzella2009}{2009}).  As in section \ref{speczsample} we
exclude any sources which are reported to have detected X-ray
counterparts or AGN emission lines by \hyperlink{Shim2011}{Shim et
  al.}  (\hyperlink{Shim2011}{2011}),
\hyperlink{Vanzella2009}{Vanzella et al.}
(\hyperlink{Vanzella2009}{2009}) and D. Stark et al. (2013).  The
excluded sources including those with the following coordinates
(03:32:29.29, $-$27:56:19.5; 03:32:44.11, $-$27:54:52.5; 12:36:42.05,
62:13:31.7; 03:32:33.77, $-$27:52:23.7).  We tabulate the measured
$[3.6]-[4.5]$ colors and coordinates of the sources we utilize in
Table~\ref{table:3}.

\begin{table*}
\centering
\begin{minipage}{140mm}

\caption{Our reference sample of spectroscopically confirmed sources
  in the redshift range $z$ $\sim$ 4.4 $-$ 5.0. The sources are
  obtained by matching the spectroscopic redshift sample listed in
  \protect\hyperlink{Vanzella2009}{Vanzella et al.}
  (\protect\hyperlink{Vanzella2009}{2009}) with the
  \protect\hyperlink{Bouwens2015}{Bouwens et al.}
  (\protect\hyperlink{Bouwens2015}{2015}) catalog.  Source IDs are as
  in the \protect\hyperlink{Bouwens2015}{Bouwens et al.}
  (\protect\hyperlink{Bouwens2015}{2015}), \protect\hyperlink{Shim2011}{Shim et al.}
  (\protect\hyperlink{Shim2011}{2011}), or \protect\hyperlink{Stark2013}{Stark et al.}
  (\protect\hyperlink{Stark2013}{2013}) catalogs.}

\begin{center}
\begin{tabular}{ l c c r  r  }

\hline
\hline
\newline
\mbox{}
\newline
ID & RA & DEC & $z_{spec}$ & [3.6] $-$ [4.5] \\
\hline
\hline
GSWV-2426242897 &03:32:42.62 &$-$27:54:28.97       &4.400   &$-$0.5 $\pm$ 0.1\\
GSDV-2228872758 &03:32:22.88 &$-$27:47:27.58       &4.440    & $-$0.3 $\pm$ 0.1\\
 GSDV-2229762901 &03:32:22.97 &$-$27:46:29.01       &4.500   &$-$0.4 $\pm$ 0.1\\
 ERSV-2285605575 &03:32:28.56 &$-$27:40:55.75       &4.597    &$-$0.2 $\pm$ 0.2\\
GSDV-2169812296 & 03:32:16.98 &$-$27:51:22.96       &4.600   &$-$0.4 $\pm$ 0.1\\
 GSWV-2475822816 &03:32:47.58 &$-$27:52:28.16       &4.758   &$-$0.7 $\pm$ 0.2\\
GSDV-2435391920 &03:32:43.53 &$-$27:49:19.20       &4.763  &$-$0.6 $\pm$ 0.1\\
GSDV-2401153550 &03:32:40.11 &$-$27:45:35.50       &4.773 & $-$0.3 $\pm$ 0.1\\
GSDV-2219353310 &03:32:21.93 &$-$27:45:33.10       &4.788   &$-$0.2 $\pm$ 0.1\\
ERSV-2052630041 & 03:32:05.26 &$-$27:43:00.41       &4.804   &$-$0.2 $\pm$ 0.1\\
GSDV-2426693897 & 03:32:42.66 &$-$27:49:38.97       &4.831  &$-$0.2 $\pm$ 0.2\\
S33166 & 03:32:58.38 & $-$27:53:39.58 &4.40 &   $-$0.3 $\pm$ 0.2\\
N12138 & 12:36:42.25 & \mbox{} 62:15:23.25 &4.414  &   $-$0.2 $\pm$ 0.1\\
N23791 & 12:37:20.58 & \mbox{} 62:11:06.11 &4.421 &   $-$0.6 $\pm$ 0.1\\
S31908 &03:32:54.04 & $-$27:50:00.81 &4.43 &   $-$0.4 $\pm$ 0.1\\
N13279 &12:36:46.16 & \mbox{} 62:07:01.83 &4.444 &   $-$0.4 $\pm$ 0.1\\
N24628 &12:37:23.57 & \mbox{} 62:20:38.72 &4.502 &    $-$0.6 $\pm$ 0.2\\
N28987 & 12:37:19.69 & \mbox{} 62:15:42.46 &4.53 &   $-$0.1 $\pm$ 0.1\\
N12849 &12:36:44.68 & \mbox{} 62:11:50.62 &4.580 &   $-$0.2 $\pm$ 0.1\\
N31130 &12:37:57.51 & \mbox{} 62:17:19.10 &4.680&   $-$0.4 $\pm$ 0.1\\
S6294 & 03:32:14.50 & $-$27:49:32.69 &4.74 &   $-$0.2 $\pm$ 0.1\\
S32900 & 03:32:57.17 & $-$27:51:45.01 &4.76 &   $-$0.5 $\pm$ 0.1\\
S1745 & 03:32:05.26 & $-$27:43:00.42 &4.80 &   $-$0.1 $\pm$ 0.1\\
S3792 & 03:32:10.03 & $-$27:41:32.65 &4.81 &   $-$0.3 $\pm$ 0.1\\
S1669 & 03:32:05.08 & $-$27:46:56.52 &4.82 &   $-$0.2 $\pm$ 0.1\\
N23039 & 12:37:18.07 & \mbox{} 62:16:41.72 &4.822 &   $-$0.3 $\pm$ 0.3\\
N6738  & 12:36:23.56 & \mbox{} 62:15:20.30 &4.889 &   $-$0.2 $\pm$ 0.2\\ 
N6333  & 12:36:21.94 & \mbox{} 62:15:17.12 &4.890 &   $-$0.2 $\pm$ 0.1\\ 
S20041 & 03:32:33.48 & $-$27:50:30.00 &4.90 &   $-$0.1 $\pm$ 0.1\\
S23745 & 03:32:44.07 & $-$27:42:27.43 &4.923 &   $-$0.5 $\pm$ 0.1\\
\hline
\end{tabular}\\
\end{center}
\label{table:3}
\end{minipage}
\end{table*}

\label{lastpage}


\begin{thebibliography}{99}
\hypertarget{Ashby2013}{\bibitem[\protect\citeauthoryear{Ashby}{2013}]{bu} Ashby M. L. N., Stanford S. A., Brodwin M., Gonzalez A. H. et al., 2013, ApJ, 769, 80}
\hypertarget{Ashby2015}{\bibitem[\protect\citeauthoryear{Ashby}{2015}]{bu1} Ashby, M.~L.~N., Willner, S.~P., Fazio, G.~G., et al.\ 2015, ApJS, 218, 33}
\hypertarget{AndersFA2003}{\bibitem[\protect\citeauthoryear{AndersFA}{2003}]{bu2} Anders P., \& Fritze-v. Alvensleben U., 2003, A\&A, 401, 1063}
\hypertarget{Bacon2015}{\bibitem[\protect\citeauthoryear{Bacon}{2015}]{bu3} Bacon, R., Brinchmann, J., Richard, J., et al.\ 2015, A\&A, 575, A75}
\hypertarget{Bouche2010}{\bibitem[\protect\citeauthoryear{Bouche}{2010}]{bu4} Bouch\'{e} N., Dekel A., Genzel R., Cresci G., F\"{o}rster Schreiber N. M., Shapiro K. L., Davies R. I., Tacconi L., 2010, ApJ, 718, 1001}
\hypertarget{Bouwens2009}{\bibitem[\protect\citeauthoryear{Bouwens}{2009}]{bu5} Bouwens R. J., Illingworth G. D., Franx M., Chary R. -R., Meurer G. R., Conselice C.J., Ford H., Giavalisco M., van Dokkum P. G., 2009, ApJ, 705, 936}
\hypertarget{Bouwens2011a}{\bibitem[\protect\citeauthoryear{Bouwens}{2011a}]{bu6} Bouwens R. J., Illingworth G. D., Labb\'{e} I., Oesch P. A., Trenti M., Carollo C. M., van Dokkum P. G., Franx M., Stiavelli M., Gonz\'{a}lez V., Magee D., Bradley L., 2011a, Nature, 469, 504}
\hypertarget{Bouwens2012b}{\bibitem[\protect\citeauthoryear{Bouwens}{2012b}]{bu7} Bouwens R. J., Illingworth G. D., Oesch P. A., Franx M., Labbe\'{e} I., Trenti M., van Dokkum P. G., Carollo C. M., Gonz\'{a}lez V., Smit R., Magee D., 2012, ApJ, 754, 83}
\hypertarget{Bouwens2014}{\bibitem[\protect\citeauthoryear{Bouwens}{2014}]{bu9} 
Bouwens, R.~J., Illingworth, G.~D., Oesch, P.~A., et al.\ 2014, ApJ, 793, 115}
\hypertarget{Bouwens2015}{\bibitem[\protect\citeauthoryear{Bouwens}{2015}]{bu10} Bouwens, R.~J., 
Illingworth, G.~D., Oesch, P.~A., et al.\ 2015, ApJ, 803, 34}
\hypertarget{Bowler2014}{\bibitem[\protect\citeauthoryear{Bowler}{2014}]{bu11}Bowler, R.~A.~A., Dunlop, J.~S., McLure, R.~J., et al.\ 2014, MNRAS, 440, 2810}
\hypertarget{Brammer2008}{\bibitem[\protect\citeauthoryear{Brammer}{2008}]{bu12} Brammer G. B., van Dokkum P. G., Coppi P., 2008, ApJ, 686, 1503}
\hypertarget{BruzualCharlot2003}{\bibitem[\protect\citeauthoryear{BruzualCharlot}{2003}]{bu13} Bruzual G., Charlot S., 2003, MNRAS, 344, 1000}
\hypertarget{Calzetti2000}{\bibitem[\protect\citeauthoryear{Calzetti2000}{2000}]{bu14} Calzetti D., Armus L., Bohlin R. C., Kinney A. L., Koornneef J., Storchi-Bergmann T., 2000, ApJ, 533, 682}
\hypertarget{Capak2015}{\bibitem[\protect\citeauthoryear{Capak2015}{2015}]{bu15} Capak, P.~L., Carilli, 
C., Jones, G., et al.\ 2015, Nature, 522, 455}
\hypertarget{Daddi2007}{\bibitem[\protect\citeauthoryear{Daddi}{2007}]{bu16} Daddi E., Dickinson M., Morrison G. et al., 2007, ApJ, 670, 156}
\hypertarget{Dave2011}{\bibitem[\protect\citeauthoryear{Dave2011}{2011}]{bu18} Dav\'{e} R., Oppenheimer B. D., Finlator K., 2011, MNRAS, 415, 11}
\hypertarget{debarros2014}{\bibitem[\protect\citeauthoryear{debarros2014}{2011}]{bu17} de Barros, S., Schaerer, D., \& Stark, D.~P.\ 2014, A\& A, 563, A81}
\hypertarget{Erb2006}{\bibitem[\protect\citeauthoryear{Erb}{2006}]{bu19} Erb D. K., Steidel C. C., Shapley A. E., Pettini M., Steidel C. C., Reddy N. A., Adelberger K. L., 2006, ApJ, 647, 128}
\hypertarget{Eyles2005}{\bibitem[\protect\citeauthoryear{Eyles}{2005}]{bu20}Eyles L. P., Bunker A. J., Stanway E. R., Lacy M., Ellis R. S., Doherty M., 2005, MNRAS, 364, 443}
\hypertarget{Finlator2007}{\bibitem[\protect\citeauthoryear{Finlator}{2007}]{bu21} Finlator K., Dav\'{e} R., Oppenheimer B. D., 2007, MNRAS, 376, 1861}
\hypertarget{Finlator2011}{\bibitem[\protect\citeauthoryear{Finlator}{2011}]{bu22} Finlator, K., Oppenheimer, B.~D., \& Dav{\'e}, R.\ 2011, MNRAS, 410, 1703}
\hypertarget{Finkelstein2012}{\bibitem[\protect\citeauthoryear{Finkelstein}{2012}]{bu122} Finkelstein, S.~L., Papovich, C., Salmon, B., et al.\ 2012, ApJ, 756, 164}
\hypertarget{Fumagalli2012}{\bibitem[\protect\citeauthoryear{Fumagalli}{2012}]{bu23} Fumagalli M., Patel S. G., Franx M., Brammer G. B., van Dokkum P. G., da Cunha E., Kriek M., Lundgren B. et al., 2012, ApJS, 757, L22}
\hypertarget{Giavalisco2004}{\bibitem[\protect\citeauthoryear{Giavalisco}{2004}]{bu24} Giavalisco, M.,  Ferguson, H.~C., Koekemoer, A.~M. et al.\ 2004a, ApJ, 600, L93}
\hypertarget{Gonzalez2010}{\bibitem[\protect\citeauthoryear{Gonzalez}{2010}]{bu25} Gonz\'{a}lez V., Labb\'{e} I., Bouwens R. J., Illingworth G. D., Franx M. et al., 2010, ApJ, 713, 115}
\hypertarget{Gonzalez2013}{\bibitem[\protect\citeauthoryear{Gonzalez}{2013}]{bu26} Gonz\'{a}lez V., Bouwens R. J., Illingworth G., Labbe\'{e} I., Oesch P., Franx M., Magee D., 2014, ApJ, 781, 34}
\hypertarget{Grogin2011}{\bibitem[\protect\citeauthoryear{Grogin}{2011}]{bu27} Grogin N. A., Kocevski D. D., Faber S. M., Ferguson H. C., Koekemoer A. M. et al., 2011, ApJS, 197, 35}
\hypertarget{Holden2015}{\bibitem[\protect\citeauthoryear{Holden}{2015}]{bu100}Holden, B.~P., Oesch, P.~A., Gonzalez, V.~G., et al.\ 2014, ApJ, submitted, arXiv:1401.5490}
\hypertarget{Kennicutt1998}{\bibitem[\protect\citeauthoryear{Kennicutt}{1998}]{bu28} Kennicutt Jr. R.~C., 1998, A\&A, 36, 189}
\hypertarget{Koekemoer2011}{\bibitem[\protect\citeauthoryear{Koekemoer}{2011}]{bu29} Koekemoer A. M., Faber S. M., Ferguson H. C., Grogin N. A., Kocevski D. D. et al., 2011, ApJS, 197, 36}
\hypertarget{Kriek2009}{\bibitem[\protect\citeauthoryear{Kriek}{2009}]{bu30} Kriek M., van Dokkum P. G., Labb\'{e} I., Franx M., Illingworth G. D., Marchesini D., Quadri R. F., 2009, ApJ, 700, 221}
\hypertarget{Kron1980}{\bibitem[\protect\citeauthoryear{Kron}{1980}]{bu31} Kron R. G., 1980, ApJS, 43, 305}
\hypertarget{Labbe2010a}{\bibitem[\protect\citeauthoryear{Labbe}{2010a}]{bu32} Labb\'{e} I., Gonz\'{a}lez V., Bouwens R. J., Illingworth G. D., Oesch P. A., van Dokkum P. G., Carollo C. M., Franx M., Stiavelli M., Trenti M., Magee D., Kriek M., 2010a, ApJL, 708, L26}
\hypertarget{Labbe2010b}{\bibitem[\protect\citeauthoryear{Labbe}{2010b}]{bu33} Labb\'{e} I., Gonz\'{a}lez V., Bouwens R. J., Illingworth G. D., Franx M., Trenti M., Oesch P. A., van Dokkum P. G., Stiavelli M., Carollo C. M., Kriek M., Magee D., 2010b, ApJL, 716, L103}
\hypertarget{Labbe2013}{\bibitem[\protect\citeauthoryear{Labbe}{2013}]{bu34} Labb\'{e} I., Oesch P. A., Bouwens R. J., Illingworth G. D., Magee D., Gonz\'{a}lez V., Carollo C. M., Franx M. et al., 2013, ApJL, 777, L19}
\hypertarget{Labbe2014}{\bibitem[\protect\citeauthoryear{Labbe}{2014}]{bu35} Labbe, I., Oesch, P., 
Illingworth, G., et al.\ 2014, Spitzer Proposal, 11134}
\hypertarget{Labbe2015}{\bibitem[\protect\citeauthoryear{Labbe}{2015}]{bu36} Labb\'{e}, I., Oesch, P.~A., Illingworth, G.~D., et al. 2015, ApJS, in press, arXiv:1507.08313}
\hypertarget{Meurer1999}{\bibitem[\protect\citeauthoryear{Meurer}{1999}]{bu37} Meurer G. R., Heckman T. M., Calzetti D., 1999, ApJ, 521, 64}
\hypertarget{NeisteinDekel2008}{\bibitem[\protect\citeauthoryear{NeisteinDekel}{2008}]{bu38} Neistein E., Dekel A., 2008, MNRAS, 388, 1792}
\hypertarget{Noeske2007}{\bibitem[\protect\citeauthoryear{Noeske}{2007}]{bu39} Noeske K. G., Weiner B. J., Faber S. M., Papovich C. et al., 2007, ApJL, 660, L43}
\hypertarget{Noll2009}{\bibitem[\protect\citeauthoryear{Noll}{2009}]{bu40} Noll, S., Burgarella, D., Giovannoli, E., et al.\ 2009, A\&A, 507, 1793}
\hypertarget{Oesch2013}{\bibitem[\protect\citeauthoryear{Oesch}{2013}]{bu41} Oesch, P.~A., Labb{\'e}, 
I., Bouwens, R.~J., et al.\ 2013, ApJ, 772, 136 }
\hypertarget{OkeGunn1983}{\bibitem[\protect\citeauthoryear{OkeGunn}{1983}]{bu42} Ok, J. B., Gunn J. E., 1983, ApJ, 266, 713}
\hypertarget{Reddy2012}{\bibitem[\protect\citeauthoryear{Reddy}{2012}]{bu43} Reddy N., Dickinson M., Elbaz D., Morrison G. et al., 2012a, ApJ, 744, 154}
\hypertarget{Reddy2012b}{\bibitem[\protect\citeauthoryear{Reddy}{2012b}]{bu44} Reddy N. A., Pettini M., Steidel C. C., Shapley A. E., Erb D. K., Law D. R., 2012b, ApJ, 754, 25}
\hypertarget{RB2015}{\bibitem[\protect\citeauthoryear{RB2015}{2015}]{bu45} Roberts-Borsani, G.~W., Bouwens, R.~J., Oesch, P.~A., et al.\ 2015, ApJ, submitted, arXiv:1506.00854}
\hypertarget{Salmon2015}{\bibitem[\protect\citeauthoryear{Salmon}{2015}]{bu46}  Salmon, B., Papovich, C., Finkelstein, S.~L., et al.\ 2015, ApJ, 799, 183 }
\hypertarget{Salpeter1955}{\bibitem[\protect\citeauthoryear{Salpeter}{1955}]{bu47} Salpeter E. E., 1955, ApJ, 121, 161}
\hypertarget{Sanders2015}{\bibitem[\protect\citeauthoryear{Sanders}{2015}]{bu48} Sanders, R.~L., Shapley, A.~E., Kriek, M., et al., 2015, ApJ, 799, 138 }
\hypertarget{SchaererdeBarros2009}{\bibitem[\protect\citeauthoryear{SchaererdeBarros}{2009}]{bu49} Schaerer D., de Barros S., 2009, A\&A, 502, 423}
\hypertarget{SchaererdeBarros2010}{\bibitem[\protect\citeauthoryear{SchaererdeBarros}{2010}]{bu50} Schaerer D., de Barros S., 2010,  A\&A, 515, A73}
\hypertarget{Schenker2013}{\bibitem[\protect\citeauthoryear{Schenker}{2013}]{bu51} Schenker M. A., Ellis R. S., Konidaris N. P., Stark D. P., 2013, ApJ, 777, 67}
\hypertarget{Shim2011}{\bibitem[\protect\citeauthoryear{Shim}{2011}]{bu52} Shim H., Chary R.-R., Dickinson M., Lin L., Spinrad H., Stern D., Yan C. -H., 2011, ApJ, 738, 69}
\hypertarget{Skelton2014}{\bibitem[\protect\citeauthoryear{Skelton}{2014}]{bu53} Skelton R. E., Whitaker K. E., Momcheva I. G. et al., 2014, ApJS, 214, 24}
\hypertarget{Smit2014}{\bibitem[\protect\citeauthoryear{Smit}{2014}]{bu54} Smit R., Bouwens R. J., Labb\'{e} I., Zheng W. et al., 2014, ApJ, 784, 58}
\hypertarget{Smit2015}{\bibitem[\protect\citeauthoryear{Smit}{2015}]{bu55} Smit, R., Bouwens, R.~J.,  Franx, M., et al.\ 2015, ApJ, 801, 122}
\hypertarget{Stark2009}{\bibitem[\protect\citeauthoryear{Stark}{2009}]{bu56} Stark D. P., Ellis R. S., Bunker A., Bundy K., Targett T., Benson A., Lacy M., 2009, ApJ, 697, 1493}
\hypertarget{Stark2010}{\bibitem[\protect\citeauthoryear{Stark}{2010}]{bu57} Stark, D.~P., Ellis, R.~S., Chiu, K., Ouchi, M., \& Bunker, A.\ 2010, MNRAS, 408, 1628}
\hypertarget{Stark2013}{\bibitem[\protect\citeauthoryear{Stark}{2013}]{bu58} Stark D. P., Schenker M. A., Ellis R., Robertson B., McLure R., Dunlop J., 2013, ApJ, 763, 129}
\hypertarget{Vanzella2009}{\bibitem[\protect\citeauthoryear{Vanzella}{2009}]{bu59} Vanzella E., Giavalisco M., Dickinson M., Cristiani S., Nonino M., Kuntschner H., Popesso P., Rosati P., Renzini A., Stern D., Cesarsky C., Ferguson H. C., Fosbury R. A. E., 2009, ApJ, 695, 1163}
\hypertarget{Weinmann2011}{\bibitem[\protect\citeauthoryear{Weinmann}{2011}]{bu60} Weinmann S. M., Neistein E., Dekel A., 2011, MNRAS, 417, 2737}
\hypertarget{Windhorst2011}{\bibitem[\protect\citeauthoryear{Windhorst}{2011}]{bu61} Windhorst R. A., Cohen S. H., Hathi N. P. et al., 2011, ApJS, 193, 27}
\hypertarget{Yabe2009}{\bibitem[\protect\citeauthoryear{Yabe}{2009}]{bu62} Yabe K., Ohta K., Iwata I., Sawicki M., Tamura N., Akiyama M., Aoki K., 2009, ApJ, 693, 507}
\hypertarget{Yan2006}{\bibitem[\protect\citeauthoryear{Yan}{2006}]{bu63} Yan H., Dickinson M., Giavalisco M., Stern D., Eisenhardt P. R. M., Ferguson H. C., 2006, ApJ, 651, 24}

\end{thebibliography}
\end{document}